\documentclass[authoryear,12pt]{elsarticle}
\usepackage{graphicx}
\def\lsim{\ ^{<}\!\!\!\!_{\sim}\>}
\def\gsim{\ ^{>}\!\!\!\!_{\sim}\>}

\begin{document}



\centerline{\large\bf Physical and dynamical properties}

\centerline{\large\bf of the anomalous comet 249P/LINEAR}

\vspace{2cm}

Julio A. Fern\'andez$^{(1,*)}$, Javier Licandro$^{(2),(3)}$, Fernando Moreno$^{(4)}$, Andrea Sosa$^{(5)}$, Antonio Cabrera-Lavers$^{(2),(6)}$, Julia de Le\'on$^{(2),(3)}$, Peter Birtwhistle$^{(7)}$

\bigskip

(1) Departamento de Astronom\'ia, Facultad de Ciencias, Universidad de la Rep\'ublica, Igu\'a 4225, 14000 Montevideo, Uruguay

(2) Instituto de Astrof\'isica de Canarias (IAC), C/V\'ia L\'actea s/n, 38205 La Laguna, Tenerife, Spain

(3) Departamento de Astrof\'{\i}sica, Universidad de La Laguna, 38206 La Laguna, Tenerife, Spain

(4) Instituto de Astrof\'\i sica de Andaluc\'\i a, CSIC, Glorieta de la Astronom\'\i a s/n, 18008 Granada, Spain

(5) PDU Ciencias F\'isicas, Centro Universitario de la Regi\'on Este (CURE), Universidad de la Rep\'ublica, 27000 Rocha, Uruguay

(6) Gran Telescopio Canarias (GTC), E-38712, Bre\~na Baja, La Palma, Spain

(7) Great Shefford Observatory, Phlox Cottage, Wantage Road, Great Shefford, Berkshire RG17 7DA, United Kingdom

\vspace{1.5cm}


\noindent Number of pages: 29\\
Number of figures: 11\\
Number of tables: 3

\vspace{1cm}

\centerline{ICARUS, in press}

\vspace{1cm}

\centerline{April 2017}

\vspace{1.5cm}

\noindent $^{(*)}$ Corresponding author\\
Phone: 5982 525 8624\\
Fax: 5982 525 0580\\
E-mail address: julio@fisica.edu.uy

\vfill
\eject

\centerline{\bf Abstract}

\bigskip

Images and low-resolution spectra of the near-Earth Jupiter family comet (JFC) 249P/LINEAR in the visible range  obtained with the instrument OSIRIS in the 10.4m Gran Telescopio Canarias (GTC) (La Palma, Spain) on January 3, 4, 6 and February 6, 2016 are presented, together with a series of images obtained with the 0.4m telescope of the Great Shefford Observatory obtained on Oct. 22 and 27, and Nov. 1 and 24, 2006. The reflectance spectrum of 249P is similar to that of a B-type asteroid. The comet has an absolute (visual) nuclear magnitude $H_V=17.0\pm 0.4$, which corresponds to a radius of about 1-1.3 km for a geometric albedo $\sim 0.04-0.07$. From the analysis of GTC images using a Monte Carlo dust tail code we find that the time of maximum  dust ejection rate was around 1.6 days before perihelion. The analysis of the dust tails during the 2006 and 2016 perihelion approaches reveals that, during both epochs, the comet repeated the same dust ejection pattern, with a similar short-lived activity period of about 20 days (FWHM) around perihelion and a dust loss rate peaking at $145 \pm 50$ kg/s. The total dust mass ejected during its last perihelion passage was $(2.5 \pm 0.9) \times 10^8$ kg, almost all this mass being emitted before the first observation of January 3, 2016. The activity onset, duration, and total ejected mass were very similar during the 2006 perihelion passage. This amount of dust mass is very low as compared with that from other active JFCs. The past orbital evolution of 249P and 100 clones were also followed over a time scale of $\sim 5 \times 10^4$ yr. The object and more than 60\% of the clones remained bound to the near-Earth region for the whole computed period, keeping its perihelion distance within the range $q \simeq 0.4-1.1$ au. The combination of photometric and spectroscopic observations and dynamical studies show that the near-Earth comet 249P/LINEAR has several peculiar features that clearly differentiate it from typical JFCs. We may be in front of a new class of near-Earth JFC whose source region is not the distant trans-neptunian population, but much closer in the asteroid belt. Therefore, 249P/LINEAR may be a near-Earth counterpart of the so-called main-belt comets or active asteroids. 

\vspace{1cm}


\vfill
\eject

\section{Introduction}

It has been largely accepted that asteroids are inactive bodies, while comets are active, observationally characterized by the display of a coma and a tail. It has also been considered that there are clear dynamical differences: while asteroidal orbits are rather stable with Tisserand parameters with respect to Jupiter $T_J > 3$, the orbits of comets are unstable with $T_J < 3$. Yet, the situation has become more complex as some near-Earth asteroids (NEAs) are found to move on cometary orbits ($T < 3$) and have close approaches to Jupiter \citep{Fern14}, while some Jupiter family comets (JFCs) move on stable asteroidal-like orbits \citep{Fern15}. A few main-belt asteroids, that were thought to be inactive, have shown activity and for this reason they have been called {\it main-belt comets} \citep{Hsie06}, also known as {\it active asteroids}.

Of course, the presence or absence of activity depends on the heliocentric distance. Objects in the deep freeze of the trans-neptunian region are inactive, disregarding of their volatile content. The closer to the Sun, the more likely the bodies will display activity if they contain volatiles. Near-Earth objects (defined as those with perihelion distances $q < 1.3$ au) are therefore the most adequate ones to study their volatile content, since if volatiles are present on or close to their surfaces, they will very likely show up through sublimation. The near-Earth region has the additional advantage of being the best sampled given its proximity to the Earth and the Sun. For instance, the sample of NEAs with diameters $>$ 1 km is estimated to be about 90\% complete \citep{Harr15}.

The discovery of some NEAs in cometary orbits raised the question about whether comets could be disguised as asteroids, in particular those with low albedos and C, D, T or P taxonomic types \citep{Deme08,Licandro08} perhaps by the buildup of insulating dust mantles \citep{Brin80}. However the ultimate fate of small-size comets (under or around 1 km) getting close to the Sun seems to be disintegration into meteoritic dust \citep{Weav01}. The survival in the Sun's vicinity over time scales of $10^4$-$10^5$ yr seems to require a hardened structure of refractory material.  

The study of the orbital histories of near-Earth JFCs (NEJFCs) revealed the presence of a sub-sample of bodies in quite stable orbits over time scales of $5 \times 10^4$ yr. \citet{Fern15} found 14 NEJFCs moving on highly or moderately asteroidal orbits. Furthermore, some seemingly inactive NEAs have shown residual, and in some cases intermittent activity as, for example, 107P/Wilson-Harrington \citep{YFernandez97}, and (3532) Don Quixote \citep{Momm14}.

The existence of near-Earth JFCs in asteroidal orbits (from now on, simply {\it asteroidal NEJFCs}) raises the question about their provenance: Do they also come from the trans-neptunian belt and by some dynamical process they have fallen into stable orbits? Or do they come from the main asteroid belt as the NEAs, thus explaining their similar dynamical characteristics? If the latter applies they could be considered as the near-Earth counterparts of the MBCs. A close-by source or a distant one may have further implications regarding the geochemical and isotopic composition of the bodies. If asteroidal NEJFCs formed closer to the Sun, their isotopic D/H ratio in water molecules could closely match that observed in chondritic meteorites and in the ocean water. Most comets seem to have a D/H ratio greater than that observed in the terrestrial water, which seems to rule them out as the main water source. The closeness of the D/H ratio (and other isotopes like $^{15}$N/$^{14}$N) with that found in chonditric meteorites points to asteroids as the main source of the terrestrial water \citep{Mart12,Alex12}. In this regard, the discovery of asteroidal JFCs becomes very relevant since they may be the residual tail of a large population of water-rich asteroids that supplied most of the water to the primitive Earth.

In this work we present new photometric, spectroscopic, and dynamical results of 249/LINEAR (hereafter 249P), namely one of the 14 asteroidal JFCs found by \citet{Fern15}. In Section 2 we study the dynamical evolution of the comet over time scales $\sim 5 \times 10^4$ yr. In Section 3 we present the spectra of both the nucleus and the tail of the comet obtained with the 10.4m  Gran Telescopio Canarias (GTC, La Palma, Spain) in Jan. 2016 and we derive their spectral properties. In Section 4 we present the images obtained with the GTC during Jan-Feb 2016, and with the 0.4m telescope of the Great Shefford Observatory (GSO) obtained in Oct-Nov 2006 together with the results of the analysis using a Monte Carlo dust tail code. The results include the activity onset and duration, total mass ejected and the determination of the nuclear size and fraction of active area. In Section 6 we discuss the results and in Section 7 we present the conclusions.

\section{Orbital characteristics and dynamical classification}

The most relevant orbital parameters are shown in Table 1. At first sight, 249P seems to have an orbit typical of near-Earth JFCs, though \citet{Fern15} found that it belongs to a peculiar subgroup of JFCs moving on stable, asteroid-like orbits and that it has probably been in an orbit similar to the current one for a long time. To quantify this time, they introduced a dynamical parameter called the {\it capture time} defined as the time needed by a given comet to decrease its perihelion distance by 1 au down to the value it had at the discovery time. For instance, typical JFCs were found to have capture times ranging from very recent ones (one or a few passages ago) up to a few $10^3$ yr. By contrast, asteroidal NEJFCs have capture times that largely exceed $10^4$ yr and in some cases they were found to have stayed in the near-Earth region for the past $5 \times 10^4$ yr. In this regard 249P was found to be among a group of 8 comets on 'highly asteroidal' orbits with a capture time close to $5 \times 10^4$ yr (see Table 2 of Fern\'andez and Sosa, 2015).

\begin{table}[h]

\centerline{Table 1: Orbital parameters of 249P/LINEAR}

\begin{center}

  \begin{tabular}{l l} \hline

Epoch & JD 2454120.5 (2007-Jan-20.0 TDB) \\   
Perihelion distance ($q$) & $0.5105849123998849 \pm 7.2696e-7$ au \\
Semimajor axis ($a$) & $2.777175367052575 \pm 1.8323e-7$ au \\
Inclination ($i$) & $8.434078867850722 \pm 2.5975e-5$ deg \\
Argument of perihelion ($\omega$) & $64.04066337480732 \pm 8.0121e-5$ deg \\ 
Longitude of the ascending node ($\Omega$) & $240.6460476470973 \pm 3.0884e-5$ deg \\  
Tisserand parameter with respect to & 2.709 \\
Jupiter ($T_J$) &  \\ \hline
\end{tabular}

\end{center}

Source: JPL Small-Body Database
\end{table}

The new passage of 249P allowed us to re-analyze its orbital evolution from the improved orbital parameters of Table 1 that were computed from astrometric data covering the period from Oct. 19, 2006 to Feb. 20, 2015, as provided by the JPL Small-Body Database Browser. The \emph{condition code} is 0, i.e. a high-quality orbit determination. Non-gravitational parameters are not determined for this comet and are very likely negligible, as inferred by its extremely low activity (see below). We used the Bulirsch-Stoer code within the MERCURY package for the orbital integrations \citep{Cham99}. We integrated the orbit of 249P in a heliocentric frame for the past $5 \times 10^4$ yr. We considered the gravitational forces of the Sun and the eight planets. The computed orbital data was stored every 1 yr. We registered all the close encounters with Jupiter at planetocentric distances smaller than 3 Hill radii ($\sim$ 3 $\times$ 0.35 au $\simeq$ 1.05 au).

\begin{figure}[h]
\resizebox{8cm}{!}{\includegraphics{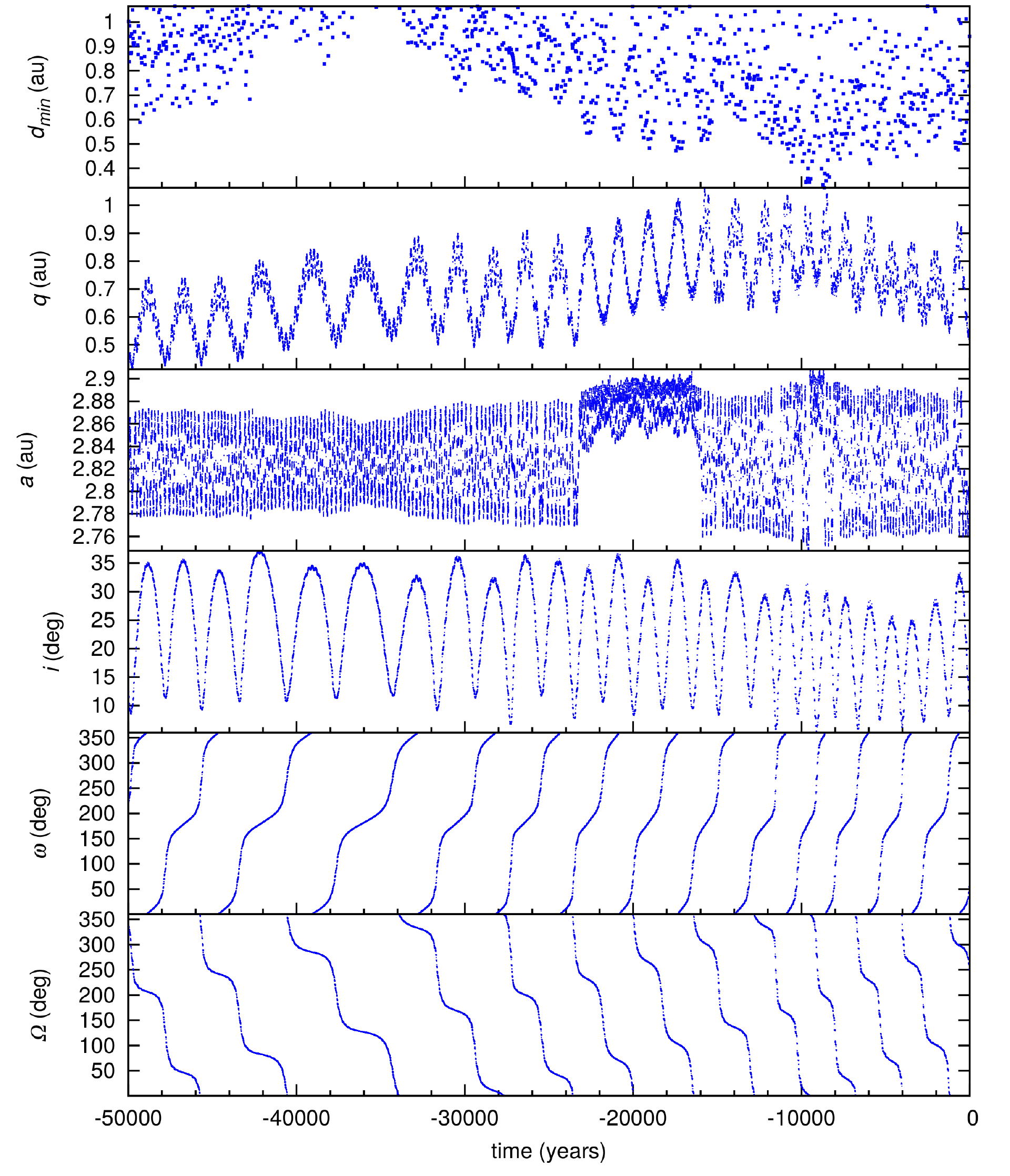}}
\caption{Orbital evolution of 249P/LINEAR for the last $5 \times 10^4$ yr. The initial orbital elements were taken from the JPL Small-Body Database Browser. We plot the following parameters: $d_{min}$: distance of closest approach to Jupiter for all encounters at less than 3 Hill radii, $q$: perihelion distance, $a$: semimajor axis, $i$: inclination, $\omega$: argument of perihelion, $\Omega$: longitude of the ascending node.}
\label{249P_evol}
\end{figure}

\begin{figure}[h]
\resizebox{10cm}{!}{\includegraphics{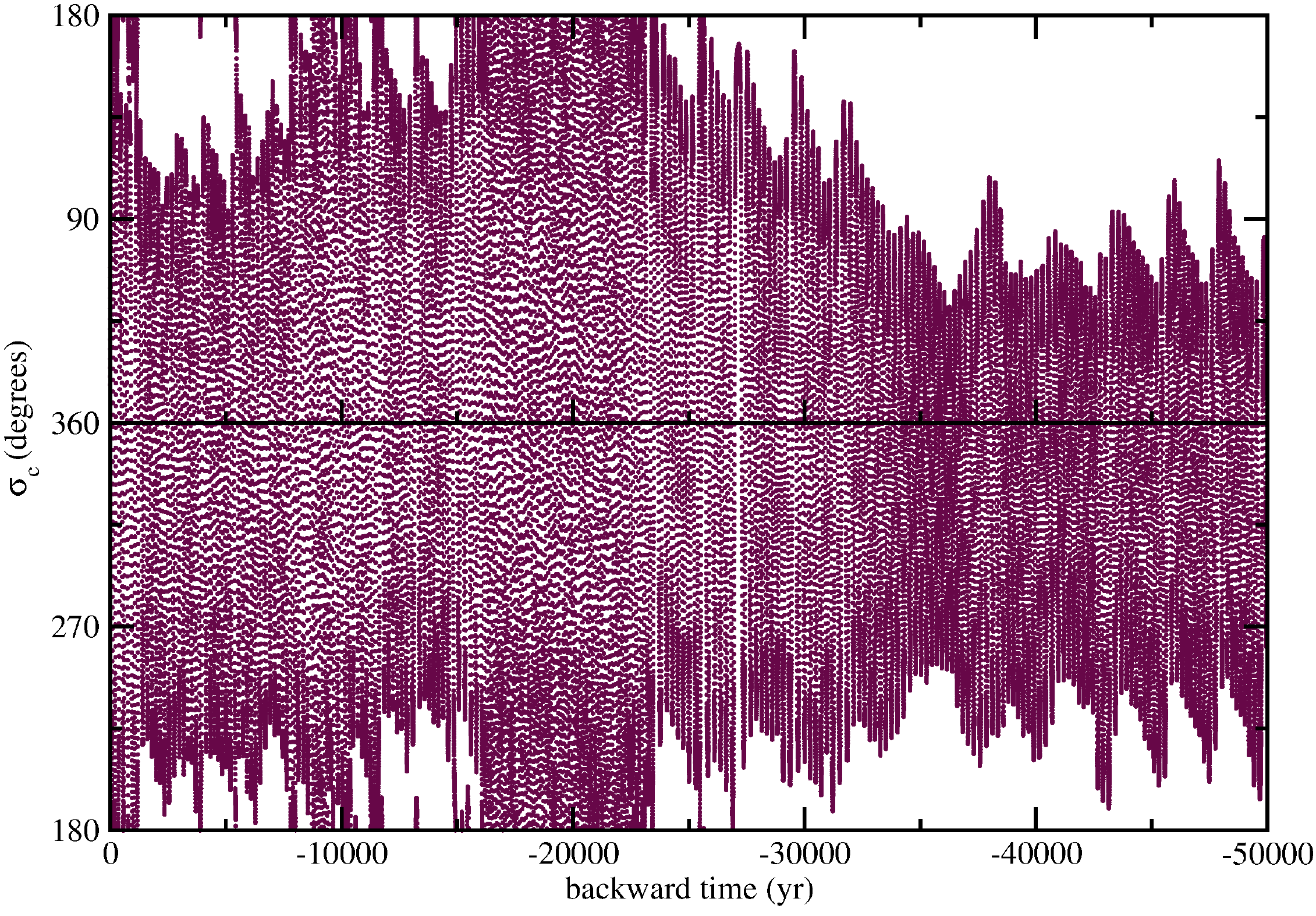}}
\caption{Libration of the critical angle of 249P around $\sigma_c = 0^{\circ}$. During part of the evolution the comet leaves the resonance and $\sigma_c$ circulates.}
\label{resonance}
\end{figure}

The results are shown in Fig. \ref{249P_evol}. The perihelion distance of the comet oscillates within $\sim 0.4-1.1$ au during the whole studied period, with a mean value $\sim 0.7$ au. The oscillations in $q$ are coupled with oscillations in $i$ (between $\sim 5-35$ degrees) due to the Kozai mechanism. 249P is found to be in the 5:2 mean motion resonance (MMR) with Jupiter for most of the computing time, with the critical angle $\sigma_c$ librating around $0^{\circ}$ (Fig. \ref{resonance}). \citet{Gall06} showed that all odd order interior resonances show librations around $\sigma_c = 0^{\circ}$ for eccentricities $e < e_c$, where $e_c = |1-[(p+k)/p]^{2/3}|$, $p$ and $k$ being integers that define the resonance (in our case, we say that it is a resonance of order $k$, with $k=3$ and $p=2$). By replacing the numerical values, the critical eccentricity is found to be $e_c=0.842$ for our odd ($k=3$) resonance, which is somewhat greater than the eccentricity of 249P ($e=0.816$), i.e. the above condition $e < e_c$ is fulfilled. The 5:2 MMR with Jupiter is one of the strongest resonances in the main asteroid belt \citep{Gall06} and several NEAs are found to be in this resonance \citep{Fern14}.  

\begin{figure}[h]
\resizebox{9cm}{!}{\includegraphics{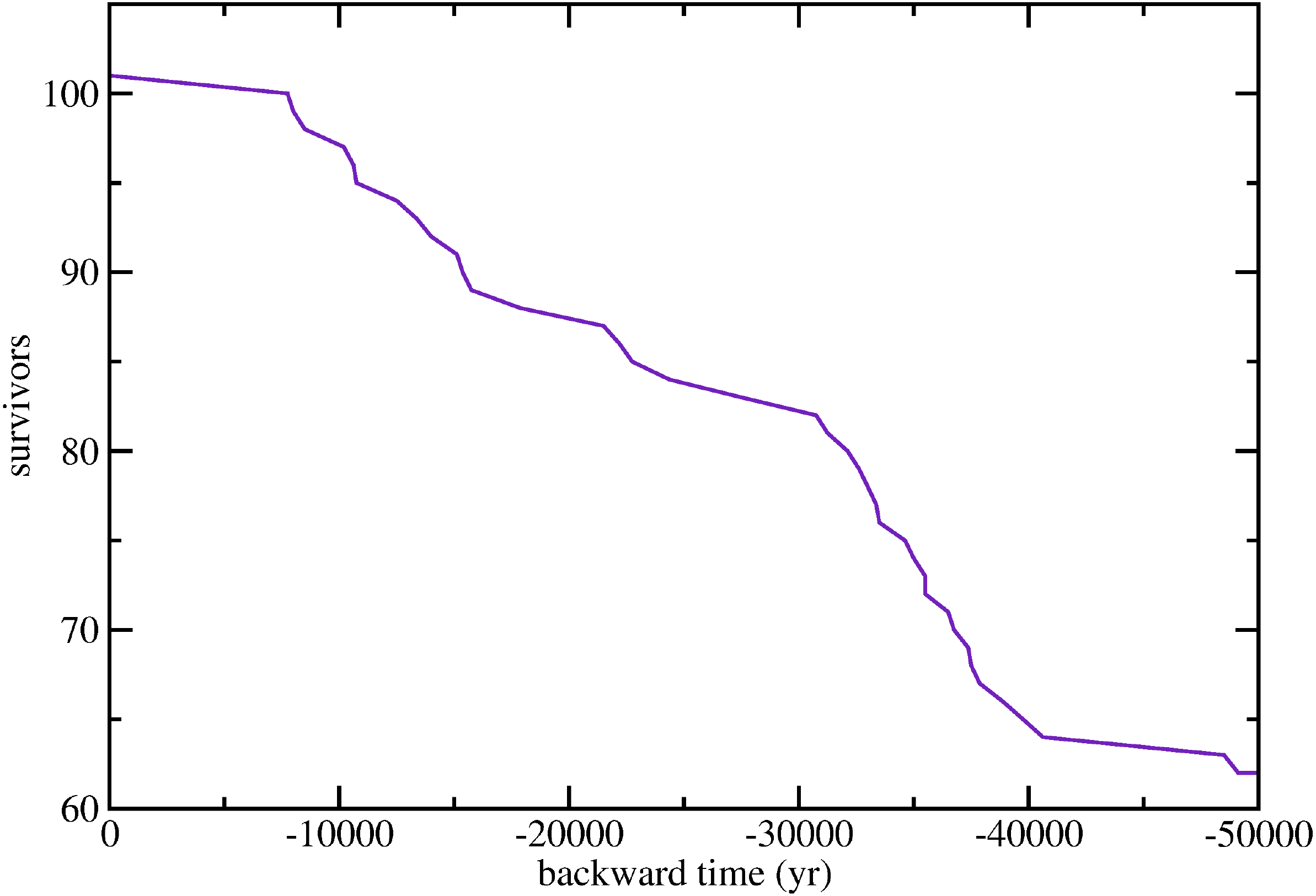}}
\caption{Clones of 249P (plus the original object) that remain in near-Earth orbit as we go backwards in time.}
\label{survivors}
\end{figure}

In order to test the robustness of the solution, we computed the orbit evolution of 100 clones of 249P by simply taking random values of the orbital parameters within Gaussian distributions, whose mean values and standard deviations were the nominal osculating values and estimated uncertainties of the comet for the given epoch (see Table 1). The comet or any of its clones was considered as \emph{ejected} if it reached a heliocentric distance greater than 100 au. Even though some of the clones get unstable orbits that move them outside the near-Earth region, the majority of them remain with perihelion distances $q < 1.5$ au for the whole studied period. Thus in the first $10^4$ yr only 3 clones are removed from the near-Earth region. Therefore, the confidence level that 249P has remained in a near-Earth orbit for the last $10^4$ yr is above 96\%. The confidence level decreases but it is still significat at -30,000 yr: it is slightly above 80\%. At the end of the studied period it is slightly above 60\% (see Fig. \ref{survivors}).

\begin{figure}[h]
\resizebox{8cm}{!}{\includegraphics{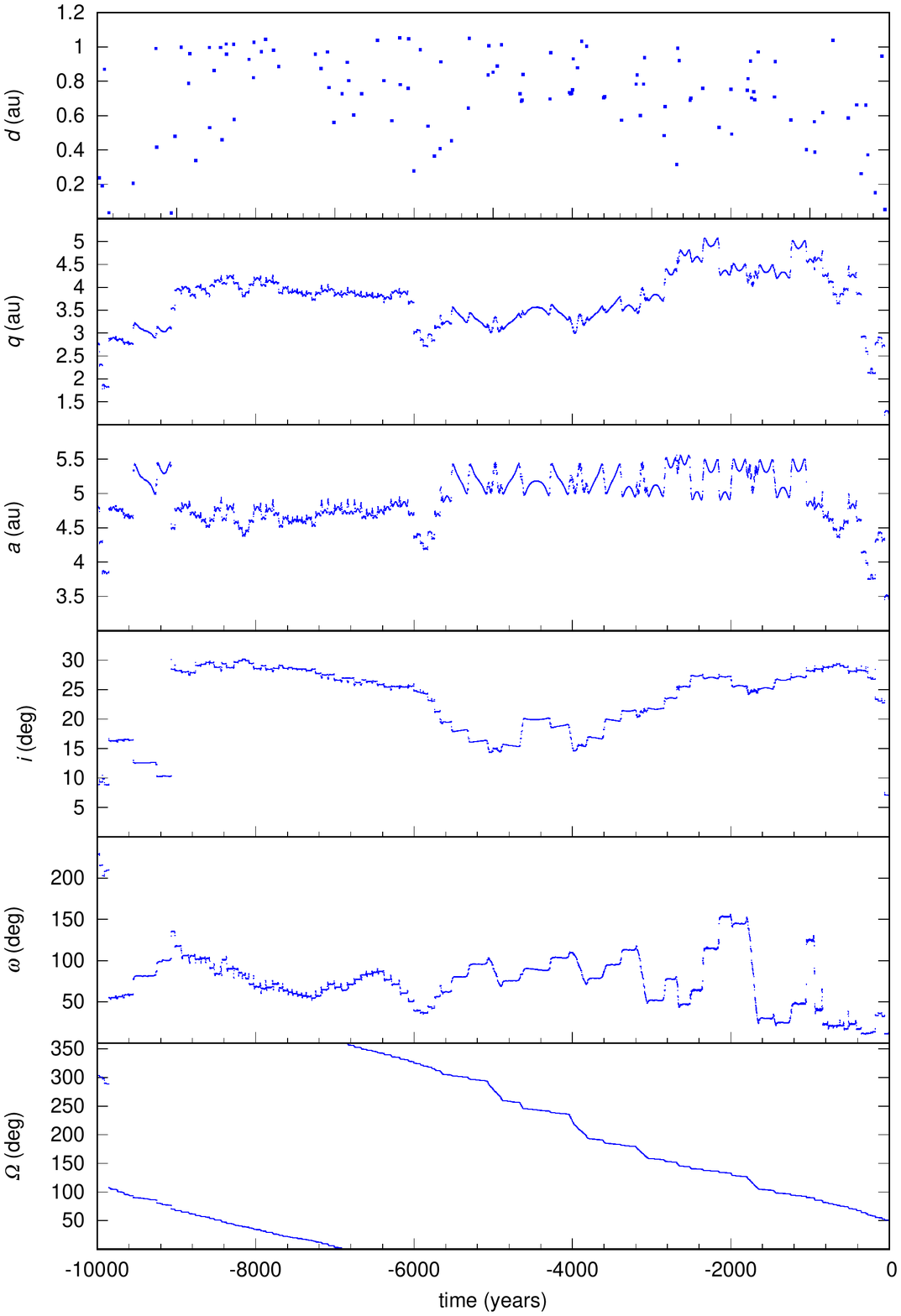}}
\caption{Orbital evolution of 67P/Churyumov-Gerasimenko for the last $10^4$ yr. The initial orbital elements were taken from the JPL Small-Body Database Browser. The parameters of the $y$-axis are the same as those shown in Fig. \ref{249P_evol}.}
\label{67P_evol}
\end{figure}

As said before, the dynamical stability of 249P sharply contrasts with the highly unstable orbits of most JFCs as expressed, e.g. by their different capture times. In order to illustrate such differences, we show in Fig. \ref{67P_evol} the case of 67P/Churyumov-Gerasimenko as an example of a typical chaotic evolution. We note that while the orbit evolution of 249P is followed through $5 \times 10^4$ yr in the past, the evolution of 67P is zoomed in for the first $10^4$ yr, which allows us to appreciate in more detail the high chaoticity of 67P's orbit.  We can see that 67P is a newcomer in the near-Earth region. It was transferred from a more distant orbit ($q=2.71$ au) to its current near-Earth orbit about 58 yr ago after a very close encounter with Jupiter (distance of closest approach $\simeq 0.0527$ au). This behavior almost repeats itself in other typical JFCs \citep{Fern15}. This is consistent with what was found by \citet{Disi09} who showed that the mean dynamical lifetime of a JFC in a Jupiter-crossing orbit is 42,300 yr, though it only spends on average about 7600 yr with $q < 1.5$ au, which indicates that it is hard for Jupiter to scatter a comet into a small-$q$ orbit (see also Fern\'andez, 1984). Therefore, the fact that 249P has spent several $10^4$ yr on an orbit with $q < 1$ au strongly suggests a different dynamics from the highly chaotic ones observed in comets coming from the trans-neptunian belt.  

\section{Spectroscopic analysis}

\subsection{Observations}

Long-slit spectra of 249P were obtained with the 10.4m GTC on January 4, 2016, 19:57hr UT at an airmass X=1.13. Two 180s spectra were done using the R300R grism (dispersion 7.74 {\AA}/pixel for the $0.6''$ slit width) of the Optical System for Image and Low Resolution Integrated Spectroscopy (OSIRIS) camera-spectrograph \citep{Cepa2000, Cepa2010}. A $2.54''$ slit width was used, oriented toward the parallactic angle to minimize slit losses due to atmospheric dispersion. 

\begin{figure}[h]
\begin{tabular}{c}
\includegraphics[width=0.5\textwidth]{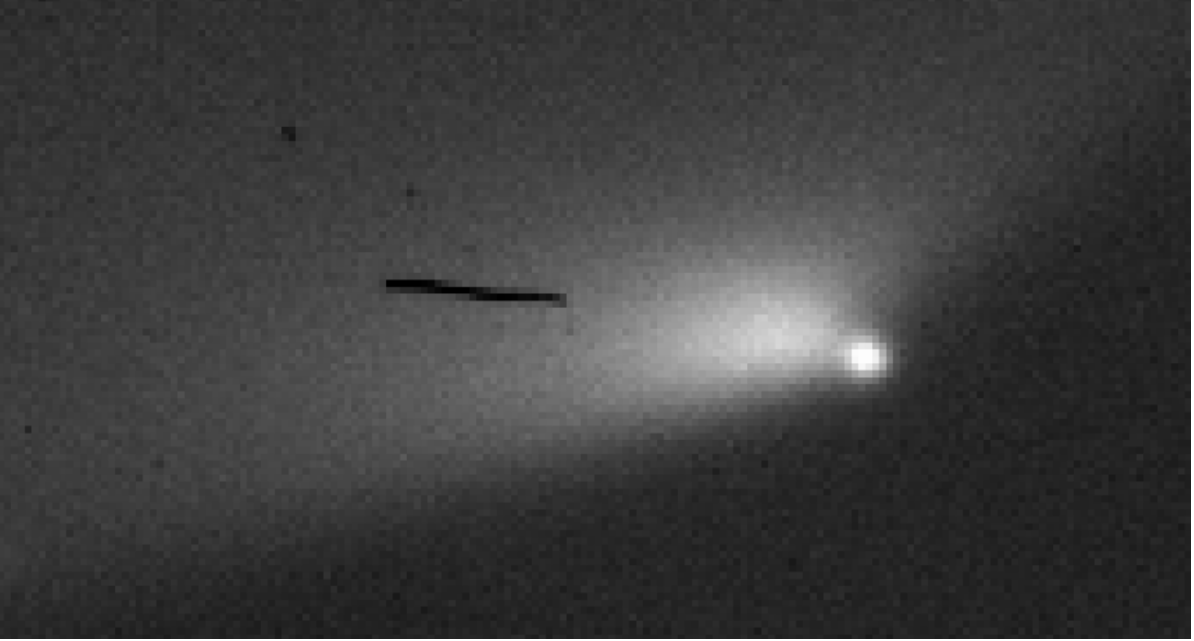}\\
\includegraphics[width=0.7\textwidth]{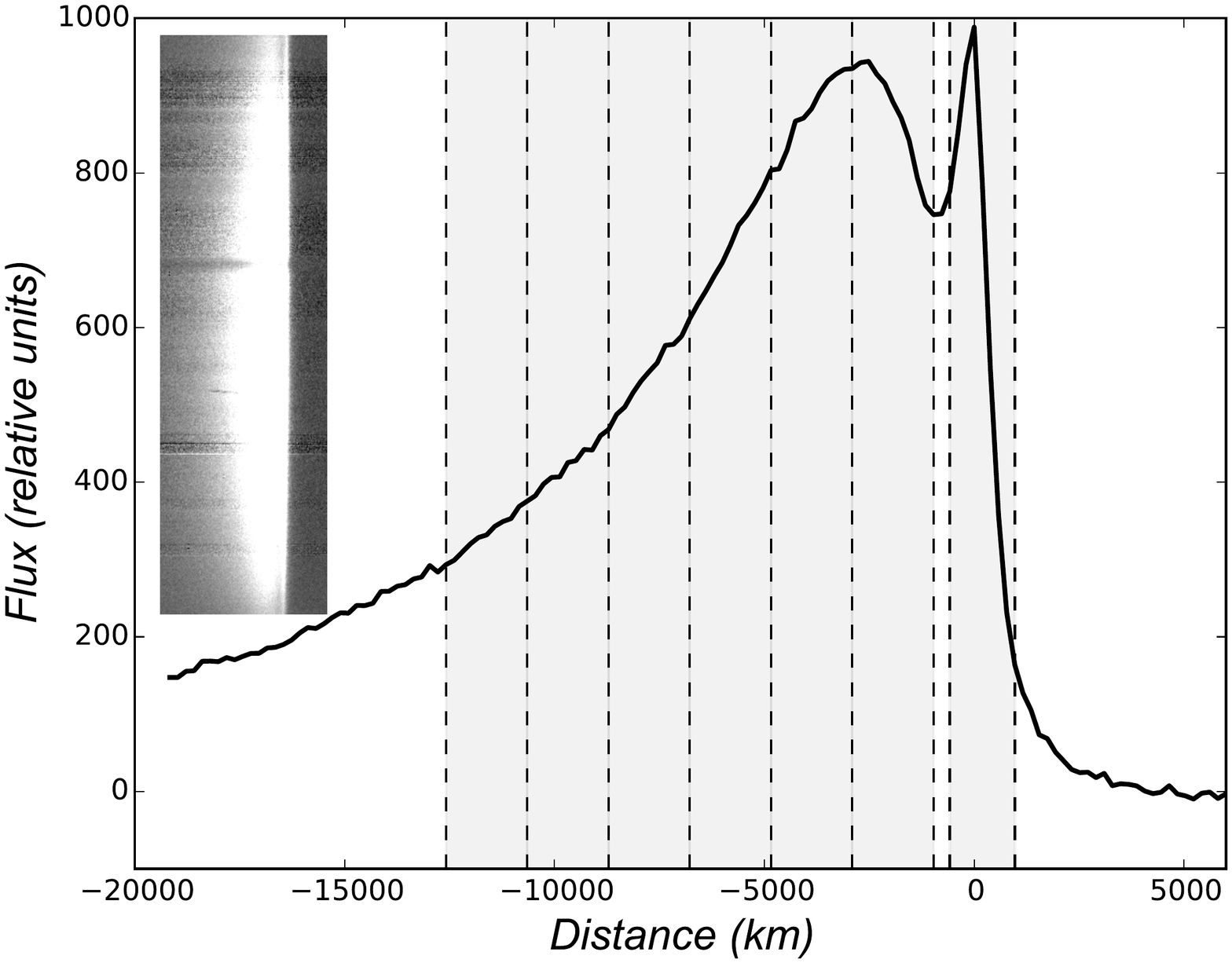}
\end{tabular}
\caption{\label{dispersion} Top panel: Image of comet 249P obtained with the 10.4m GTC telescope oriented in parallactic angles, the slit is  horizontally oriented and centered in the comet nucleus. Bottom panel: The plot is the average of the 50 central lines of the spectrum in the spatial dispersion axis. The  narrow peak centered at 0km corresponds to the nuclear region of the  comet, and the coma is to the left. The grey vertical regions limited  by dashed lines are the different apertures used to extract the  one-dimensional spectra of the nucleus and coma (see text). In the  image inside the plot is the 2-D spectrum image of comet 249P obtained  with the 10.4m GTC telescope, the x-axis is the spatial dispersion,  the y-axis is the spectra dispersion.}
\end{figure}

Images  of  the  spectra  were  bias-corrected using bias frames  and  flat-field-corrected using lamp flats. The two-dimensional spectra were wavelength-calibrated  using spectral images of the  Xe+Ne+HgAr lamps.  A profile of the observed flux in the spatial direction, obtained by averaging 50 lines of the 2-D spectrum, together with an image of the comet as seen in the slit are presented in Fig. \ref{dispersion}. In this figure the region of the nucleus corresponds to the narrow peak at 0 km, the tail is extended to the left. The FWHM of the optocenter is $\sim 1.1''$, almost the same value of the FHWM of the point-like sources observed that night, suggesting that the optocenter flux is dominated by reflected light from the nucleus. The sky background  was  subtracted by fitting the sky in the region without coma. A first-order constant sky level is fitted using a section across the spatial dispersion in the anti-tail direction (to the right of the nucleus in Fig. \ref{dispersion}) in a region close enough to the nucleus of 249P carefully avoiding any contribution from the observed tail. 

\begin{figure}[h]
  \centering
  \includegraphics[width=8.0cm]{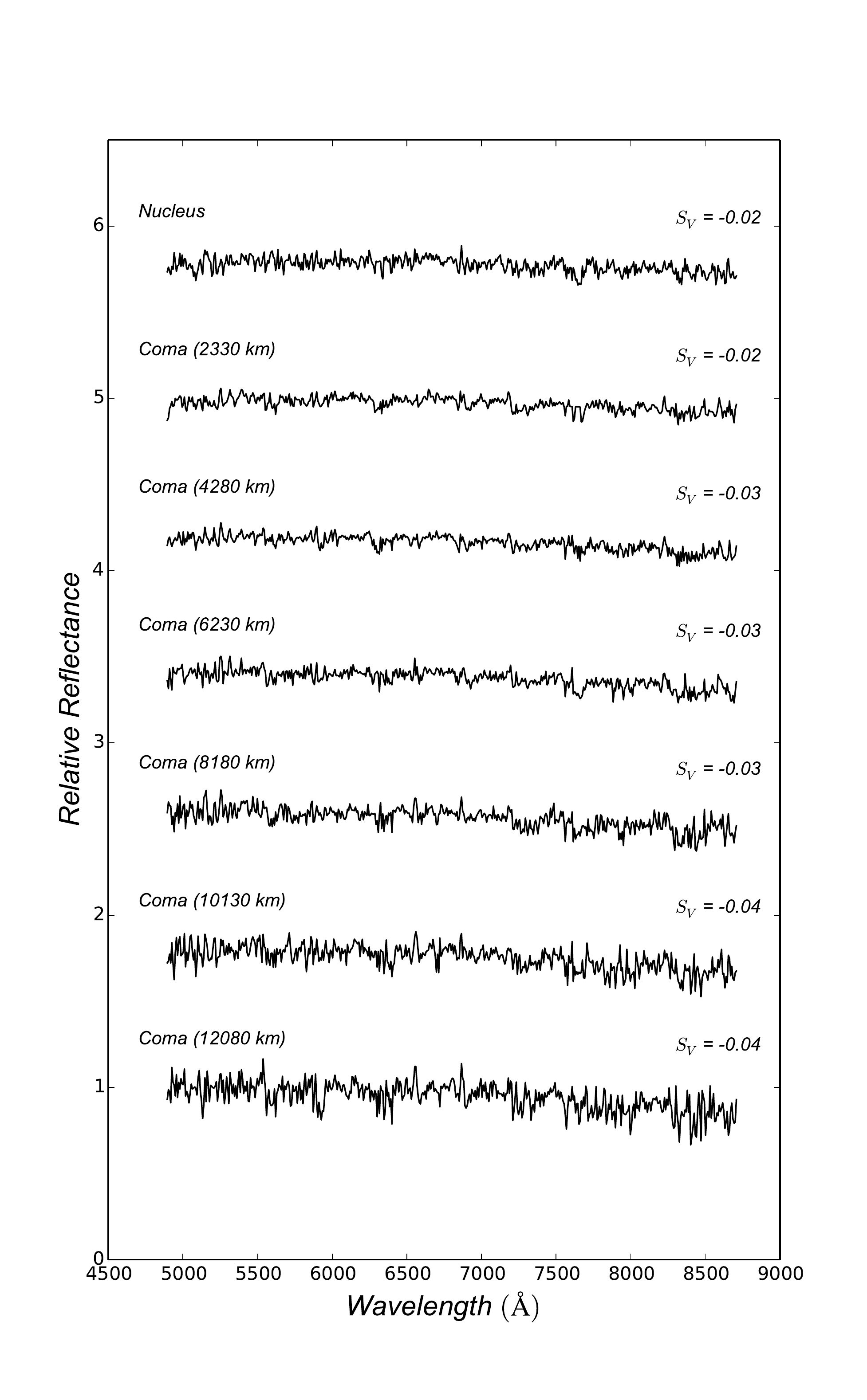}\\
  \caption{\label{espectros} Spectra of the nuclear region (labeled as {\it Nucleus}) and the coma obtained at different distances from the nucleus as indicated in the labels. All spectra are featureless with a bluish color, with spectral slope $S_V$ betwen -2.0 and -4.0$\pm 0.01$\%/1000 \AA. }
\end{figure}

The one-dimensional spectrum of the optocenter of 249P was extracted using a very narrow extraction aperture of only -3,+5 pixels (see Fig. \ref{dispersion}) to minimize the coma contribution. Additionally, six one-dimensional spectra of the coma and tail were extracted at different distances from the nucleus ($\rho$) using apertures of $\pm$ 5 pixels (see also Fig. \ref{dispersion}). To correct for telluric absorption and to  obtain  the  relative  reflectance, the G2 star Land115-271 from  the \citet{Landolt} catalogue was observed the same night immediately after the comet at 20:15 UT  using the same spectral configuration at an airmass X=1.80, and was used as a solar analogue star. The extracted spectra of 249P nucleus and coma were then divided by the corresponding extracted spectrum of the solar analog and then normalized to unity at 0.55$\mu$m. As the comet and Landolt star were observed at a slightly different airmasses, an extinction correction using the mean extinction coefficients of the ORM observatory \footnote{see http://www.ing.iac.es/Astronomy/observing/conditions/wlext.html} was used to obtain the final relative reflectance of nucleus and coma of 249P (see Fig. \ref{espectros}). Notice in any case that this is a small correction that corresponds to $\sim$ -1\%/1000 {\AA} of the final spectral slope. 

\subsection{Spectral classification \label{sec:results}}

The reflectance spectrum of the comet nuclear region (labeled as {\it Nucleus} in Fig. \ref{espectros}) and the spectra of the coma at different cometocentric distances are featureless and present a slightly blue color. The spectral slope ($S_V$) of the nucleus, computed by linear fitting the reflectance in the 0.5-0.85 $\mu$m region, is $S_V$=-2.0$\pm$1.0\%/1000 \AA, equivalent to the spectrum of a B-type asteroid in the Bus \& Binzel taxonomy \citep{Bus2002a, Bus2002b}. The spectrum of the coma is almost equal to that of the nucleus in the region close to it, with the same $S_V$, and gets slightly bluer as we move further from it up to  $S_V$=-4.0$\pm$1.0\%/1000 \AA~at $\rho =12.000$ km (see Fig. \ref{espectros}). The color of the dust coma is bluer at larger nucleocentric distances probably because of the larger abundance of smaller particles there than closer to the nucleus. Smaller particles are ejected with larger velocities than the larger ones and are accelerated more due to the solar radiation pressure. 

As discussed above, the spectrum of the optocenter of the comet is not very much contaminated by the dust coma because its FWHM is almost the same of that of the stars. To subtract the coma contribution is not trivial, but we can roughly estimate its contribution to the total flux in the selected aperture. In the images created with the model (see Section 4.2), the coma contribution in the position of the optocenter is $< 4\%$, so we can assume that the brightness profile at cometocentic distances $\rho > 0$ km to the right in Fig. \ref{dispersion}, is almost entirely due to the comet nucleus. As shown by \citet{Tanc00}, cometary comae are optically thin in little active comets (like 249P), so we do not expect any significant degree of obscuration of the nucleus, even considering unaccounted large grains given their low relative number and poor light scattering efficiency. Assuming that the brightness profile of the nucleus is symmetrical we estimate that it contributes with ~86\% of the flux in the slit. Anyhow, the contribution of the coma could change the color of the bare nucleus, but considering that the observed spectrum of the coma is almost identical to the spectrum of the optocenter, we conclude that this coma contamination has no significant effect, and the spectrum of the optocenter is likely the spectrum of the comet bare nucleus.

If the measured spectrum reflects the properties of the bare nucleus, then we have to conclude that this is a very peculiar one for a comet. A bluish to almost neutral spectral slope is atypical in cometary nuclei \citep{kolokolova2004, Licandroetal2007}. The large majority of comet nuclei correspond to P- or D-type asteroids ($S_V >$ 2\%/1000 \AA). On the other hand, a featureless with a neutral to bluish spectrum like a B- or C-type asteroid is typical of MBCs with an activity likely produced by water-ice sublimation like 133P/Elst-Pizarro, 176P/LINEAR and (300163) 2006 VW139 \citep{Licandro2011, Licandro2013}. The B-like spectral type of 249P argues against an origin in the trans-neptunian belt.

\begin{figure}[h]
  \centering
  \includegraphics[width=8.5cm]{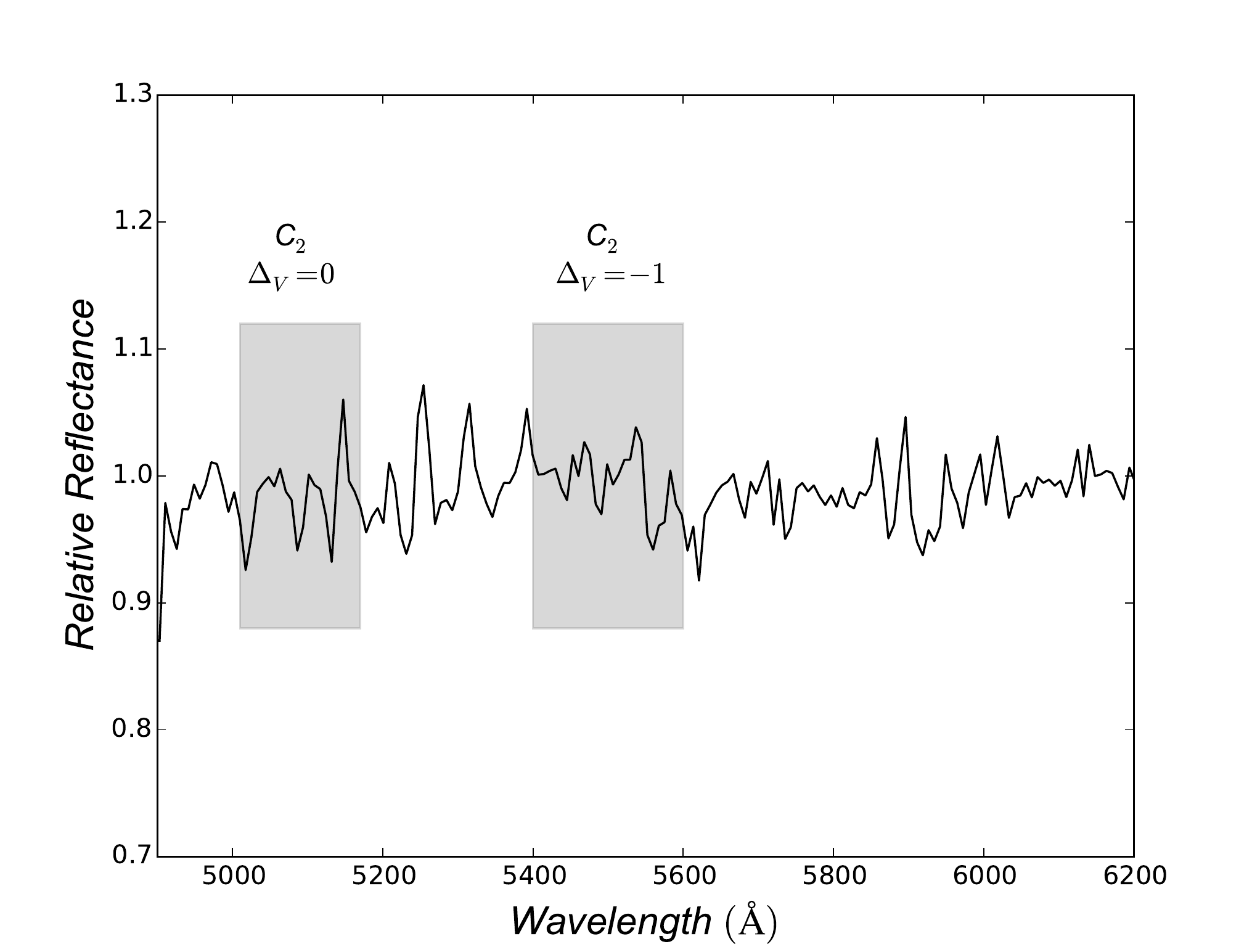}\\
  \caption{\label{C2region} Spectrum of the coma obtained in the region between 2000-7500 km from the nucleus in the region where the most prominent gas emission, that of C$_2$ Swan system, produces two prominent emission bands in comets (shown in the figure by the grey rectangles). No C$_2$ emission is detected within the SNR.}
\end{figure}

We also looked for gas emission in the coma. In the 5000-8500 {\AA} region the most prominent gas emission is that of C$_2$ Swan system that produces two prominent emission bands, one at $\sim 5100$ \AA~($\Delta_V=0$) and another at $\sim 5500$ \AA~($\Delta_V=1$). To look for them we averaged the spectra of the coma in the region between 2000-7500 km from the nucleus (see Fig. \ref{C2region}). No C$_2$ emission is detected, which is compatible with the results of the Monte Carlo coma model presented in Section 4.2 that suggests that the comet activity nearly stopped some days before the spectrum was obtained. We determined an upper limit of the C$_2$ production rate as we did in \citet{Licandro2011}. We first calibrated the spectrum in intensity (erg s$^{-1}$ cm$^{-2}$ \AA ~ster) and in projected distance (km) along the slit using the spectrophotometric standard G191-B2B observed that night. To separate the contribution of the dust we fitted the observed continuum using the solar spectrum degraded to the resolution of our observations, normalized to the flux of the comet at 5000 \AA , and subtracted it from the observed spectrum. As no gas emission is detected we determined an upper limit of the C$_2$  ($\Delta_V=0$) band measuring the standard deviation ($\sigma$) of the spectrum in a distance close to the nucleus, at  2300 km using a width of 2000 km multiplied by 33 \AA. We obtained $\sigma = 0.9 \times 10^{16}$  erg s$^{-1}$ cm$^{-2}$ \AA ~ster. Assuming that the upper limit detection of the band intensity is  3$\sigma$, and a C$_2$ g-factor of $4.5 \times 10^{-13}$ we get a upper limit for the C$_2$ column density of $\sim 5 \times 10^7$ cm$^{-2}$. Using the vectorial model by \citet{Festou1981} we computed the upper limit of C$_2$ production rate of Q(C$_2$) $\sim 1.5 \times 10^{22}$ mol/s. This value turns out to be very low as compared with C$_2$ production rates observed in active JFCs as, for instance 21P/Giacobini-Zinner, a comet of similar size as 249P \citep{Fern05}, for which \citet{Coch87} obtained Q(C$_2$) $\sim 3.3 \times 10^{24}$ mol/s at $r=1.06$ au, i.e. two orders of magnitude greater than the upper limit found for 249P at a comparable heliocentric distance. This is consistent with what we found for the dust production rate. 

\section{The nucleus, dust coma and tail. Dust production}

\subsection{Observations and data reduction}

\begin{figure}[ht]
\centerline{\includegraphics[scale=0.55,angle=-90]{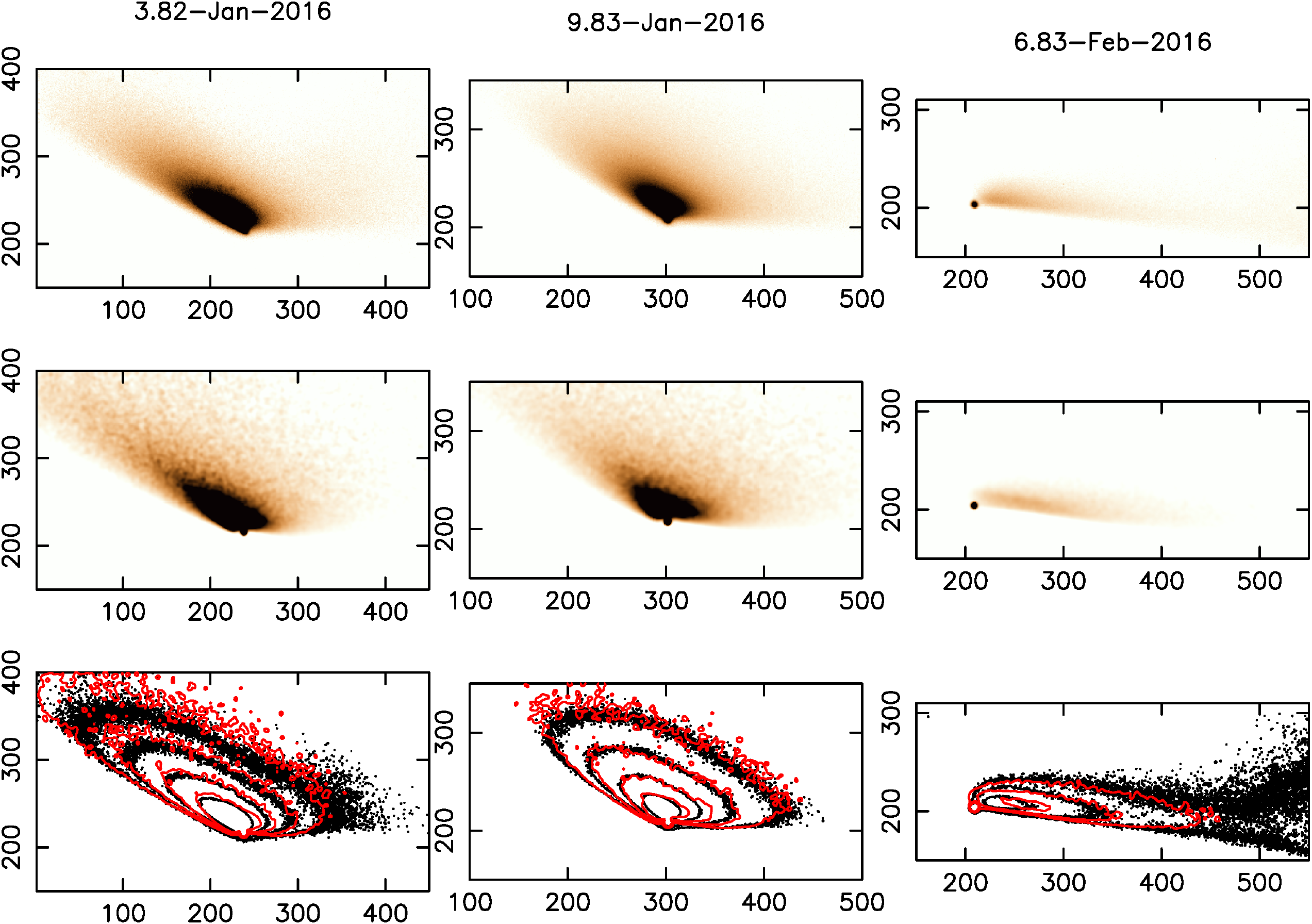}}
\caption{Upper panels: observations of comet 249P with OSIRIS at GTC
  at the indicated epochs. Middle panels: best-fit synthetic images
  obtained through minimization in the five-dimensional dust parameter
  space with the Monte Carlo dust tail code. Lower panels: isophote
  fields (black contours, observation; red contours, model). The
  innermost contour level corresponds to 10$^{-13}$ (January images)
  and 2.5$\times$10$^{-14}$ (February image) solar disk
  intensity units. Contours decrease in factors of two outwards. All the images are oriented in the North-⁠up East-⁠to-⁠the-⁠left conventional orientation. The $x$-⁠ and $y$-⁠axis are given in pixels. The physical dimensions of the images are, in km$\times$km, and from left to right, 87120$\times$48400, 81000$\times$40500, and 112080$\times$44832.}
   \label{dust_tail_1}
\end{figure}

Images of comet 249P/LINEAR were acquired under photometric and good seeing
conditions on the nights of January 3 and 6, and February 6, 2016, using a Sloan $r^\prime$ filter with OSIRIS at the 10.4m GTC telescope. The seeing disk was $1.1''$ on those nights. The image scale was $0.254''$ px$^{-1}$. The images were bias subtracted, flat-fielded, and calibrated using standard stars. A median stack image was produced each observing night from the available frames. These images were converted from mag arcsec$^{-2}$ to solar disk intensity units (the output of our Monte Carlo dust tail code) by setting $r_\odot^\prime$=--26.95, obtained assuming $V_{\odot}$=--26.75 and $(B-V)_{\odot}$=0.65 \citep{Cox00}, and the photometric relations from \citet{Fukugita96}. The log of the observations is presented in Table 2. This table includes the date of the observations (in UT and in days to perihelion), the heliocentric ($r$) and geocentric ($\Delta$) distances, the phase angle ($\alpha$), and the angle between the comet nucleus-to-Earth vector and the comet orbital plane (PlAng). The reduced images are shown in Fig. \ref{dust_tail_1}, in the conventional North-up, East-to-the-left orientation. 

\begin{figure}[ht]
\centerline{\includegraphics[scale=0.55,angle=-90]{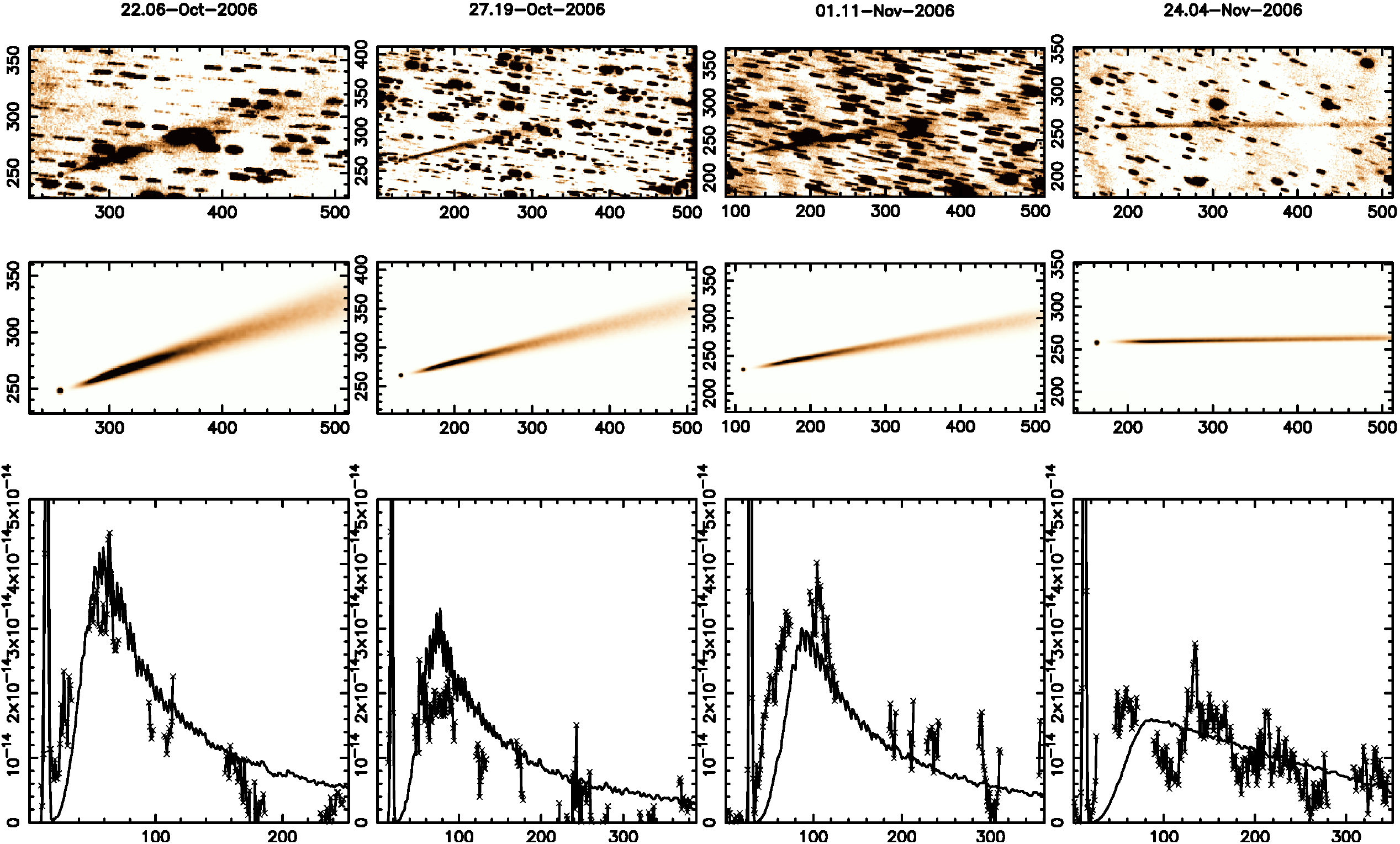}}
\caption{Upper panels: observations of comet 249P with the 0.4m telescope of
Great Shefford Observatory near 2006 perihelion at four epochs.
  Middle panels: best-fit synthetic images obtained with the same dust
  physical parameters obtained for the 2016 images, except for a
  smaller production rate. All the images are oriented in the North-⁠up
East-⁠to-⁠the-⁠left conventional orientation. The $x$-⁠ and $y$-⁠axis are given
in pixels. The physical dimensions of the images are, in km$\times$km, and from left to right, 176734$\times$83683, 266040$\times$126224,
131879$\times$133907, and 350470$\times$164982. Lower panels:
  scans along the tail for the observed (crosses joined by straight lines) and model (solid lines). The observed scans drawn are limited to the less contaminated regions by background trailed stars. The $x$-⁠ and $y$-⁠axis of the scans are expressed in pixel and solar disk intensity units, respectively.}
  \label{dust_tail_2}
\end{figure}

In addition to the GTC observations, for completeness, and in order to test the model on a previous perihelion passage in 2006, a set of reduced images from GSO (code J95) were also analyzed. These images were obtained through a clear filter on a CCD camera at the 0.4m f/6.0 Schmidt-Cassegrain telescope. The 2006 magnitudes were measured using Astrometrica software (version 4.8.2.405) and the UCAC4 star catalogue. The photometric reduction of the unfiltered images was made selecting Sloan $r'$ magnitudes from the star catalogue (that version of Astrometrica actually extracts APASS $r'$ magnitude data from the UCAC4 catalogue). Therefore the 2006 as well as the 2016 magnitudes were both calibrated directly to the Sloan $r'$ filter. As with the GTC images, the images were converted from mag arcsec$^{-2}$ to solar disk units. The image scale was 2.1 arcsec px$^{-1}$, and the field of view $18'\times 18'$. The circumstances of these observations are given in Table 2, and the images are displayed in Fig. \ref{dust_tail_2} at the four dates of observation, in the conventional North-up, East-to-the-left orientation.


\begin{table}[h]

\centerline{Table 2: Log of the observations}

\begin{center}

  \begin{tabular}{l l l l l l l l} \hline
    UT & Time to  & $r$ & $\Delta$ & Phase & PlAng & Observed ($r'$) & Synthetic ($r'$) \\
    YYYY/MM/DD & perihelion & (au) & (au) & angle ($^\circ$) & ($^\circ$)(*) & Magnitude & Magnitude \\
   HH:MM & (days) & & & & & & \\ \hline
2016/01/03 19:45 & +38.1 & 0.916 & 1.051 & 59.53 & +5.35 & 18.75$\pm$0.02 & 18.98 \\
2016/01/09 19:56 & +44.1 & 1.005 & 1.099 & 55.54 & +5.67 & 19.10$\pm$0.02 & 19.19 \\ 
2016/02/06 19:56 & +72.1 & 1.399 & 1.521 & 39.20 & +5.28 & 20.00$\pm$0.03 & 19.93 \\
2006/10/22 01:24 & +54.5 & 1.153 & 0.410 & 57.59 & --10.94 & 17.28$\pm$0.02 & 17.54 \\
2006/10/27 04:27 & +59.7 & 1.225 & 0.425 & 48.27 & --8.98 & 17.05$\pm$0.02 & 17.38 \\
2006/11/01 02:39 & +64.6 & 1.293 & 0.444 & 39.69 & --7.12 & 17.03$\pm$0.01 & 17.25 \\
2006/11/24 00:56 & +87.5 & 1.593 & 0.612 & 6.46  &  +0.20 & 16.84$\pm$0.01 & 17.07 \\ \hline
\end{tabular}
(*) Angle between the comet nucleus-to-Earth vector and the comet orbital plane.
\end{center}

\end{table}



In both data sets, 2006 and 2016, the central condensation (nucleus) appears almost separated from the tail, especially in the 2006 observations (see Figs. \ref{dust_tail_1} and \ref{dust_tail_2}). This indicates that the cometary activity has either ceased or it is very low at the time of the observations.  All together, the observations cover a wide range in the height of the Earth above the comet orbital plane: from PlAng$\sim$--11$^\circ$ to PlAng=+5.7$^\circ$, being nearly zero (Earth located on the orbital plane) during the observation of November 24, 2006, in which the tail appears as a narrow linear structure in the image.

\subsection{The Monte Carlo Dust Tail Model}\label{model}

To perform a theoretical interpretation of the obtained images in terms of the dust physical parameters, we used our Monte Carlo dust tail code, which has been used previously on several works on activated asteroids and comets, including comet 67P/Churyumov-Gerasimenko, the Rosetta target \citep[e.g.,][]{Moreno16}. We first applied the model to the GTC observations, and then we checked the applicability of the obtained dust parameters to the 2006 near-perihelion images. 

This model computes the dust tail brightness of a comet or activated asteroid by adding up the contribution to the brightness of each particle ejected from the parent nucleus. The particles, after leaving the object's surface, are ejected to space experiencing solar gravity and radiation pressure forces. The nucleus gravity force is neglected, a valid approximation for small-sized objects. Then, the trajectories of the particles become Keplerian, having orbital elements which depend on their physical properties and ejection velocities \citep[e.g.][]{Fulle89}. 

The  ratio of radiation pressure to the gravity forces exerted on each particle is given by the parameter $\beta =C_{pr}Q_{pr}/(2\rho_d a)$, where $C_{pr}$=1.19$\times$ 10$^{-3}$ kg m$^{-2}$, $Q_{pr}$ is the radiation pressure coefficient, $\rho_d$ the dust particle density and $a$ the particle radius. $Q_{pr}$ is taken as 1, as it converges to that value for absorbing spherical particles of radius $a \gsim$1 $\mu$m \citep[see e.g.][their Figure 5]{Moreno12}. 

To make the problem tractable, a number of simplifying assumptions on the dust physical parameters must be made. Thus, the particle density is taken as 1000 kg m$^{-3}$, and the geometric albedo is set to $p_v=0.04$, the same value adopted in principle for the comet nucleus, indicative of dark material of carbonaceous composition \citep[see e.g.][]{Moreno12}. For a given size distribution function, higher values of $p_v$, as a B taxonomic type might suggest, would imply a smaller dust mass loss rate in the same proportion as the albedoes to obtain the same result on the synthetic image values. On the other hand, an increase in particle density from comet-like to asteroid-like values in a given factor would require a decrease in the values of both the minimum and maximum particle radii in the same factor to maintain the same values in the $\beta$ parameter space and, in this way, to maintain the same brightness values in the synthetic images.  


For the particle phase function correction, instead of following a simple linear law (valid for small phase angles $\alpha \lsim 30^{\circ}$), we have followed the combined phase function computed by D. Schleicher \footnote{see http://asteroid.lowell.edu/comet/dustphase\_details.html} from observations of comets at all phase angles. The reason is that most of the observations were acquired at large phase angles (see Table 2), out of the region  of $\alpha \lsim 30^{\circ}$, so that the application of a simple linear phase coefficient to correct the phase function is no longer valid.

A broad size distribution is assumed, with minimum and maximum particle radii set initially to 1 $\mu$m and 1 cm, respectively, and following a power-law function of index $\kappa$=--3.5. This index was found appropriate after repeated experimentation with the code, and it is within the range of previous estimates of the size distribution of comet dust \citep[see e.g.][]{Sitk11, Ishi15, Full16}. To simplify, isotropic ejection of the particles will be assumed.

The ejection velocity of the particles depends on the activation mechanism involved. As will be shown below, the comet repeats the same activity cycle during two perihelion approaches, which is a convincing indication of typical cometary activity driven by sublimation of ices. Then, we adopt a canonical $v=v_0 \beta^{1/2}$, derived from theoretical calculations of dust drag by sublimating gases \citep[see e.g.][]{Whipple51}.

The dust mass loss rate as a function of the heliocentric distance must be specified. To model this function, we adopt a Gaussian function defined by three parameters: the peak dust mass loss rate, ($\dot{M}_0$), the time of the maximum ejection rate ($t_0$), and the full-width at half-maximum (FWHM), related to the duration of the activity.  

Summarizing, we are left with a total of four fitting parameters: $v_0$, $t_0$, $\dot{M}_0$, and FWHM. To find the best-fit set of parameters, we use the downhill simplex method \citep{Nelder65}, using the FORTRAN implementation described in \citet{Press92}. The figure-of-merit function is defined by the parameter $\chi=\sum \sigma_i$, where the summation is extended to the three dust tail images obtained, and $ \sigma_i=\sqrt{(\sum[\log(I_{obs}(i))-\log(I_{fit}(i))]^2/N(i))}$, where $I_{obs}(i)$  and $I_{fit}(i)$ are the observed and modeled tail brightness, and $N(i)$ is the number of pixels of image $i$. 

A preliminary, zero-th order, analysis of the cometary activity onset and duration is provided by the syndyne-synchrone maps \citep{Finson68}. This serves to roughly estimate the timing parameters ($t_0$ and FWHM) of the starting simplex. The rest of parameters are allowed to vary between physically reasonable limits. Thus, the starting simplex is defined with maximum values of $\dot{M}_0$=1000 kg s$^{-1}$, and $v_0$=500 m s$^{-1}$.

The brightness of the nucleus is apparent in all the images, so that it has been included in the dust tail model. For that purpose, we assume a spherical nucleus having a certain radius, with a geometric albedo of 0.04 and a linear phase coefficient of 0.04  mag deg$^{-1}$. If the albedo were larger, say $p_v = 0.07$, more in line with that expected for a B-type object (see discussion below), the modeled dust production rate and total ejected dust mass will change by a factor $\sim 4/7$. At each model iteration, the computed tail+nucleus synthetic images were convolved with the corresponding seeing disk of the observation before computing the  $\sigma_i$ values. The value of the nucleus radius does not enter as a free parameter of the model, however. We imposed several radii values around the most likely one, which will be discussed in Section 4.4.  

\subsection{Dust production near perihelion}

Using the procedure described in the previous section and assuming an albedo for the dust particles of 0.04, we found the following best-fit parameters: $\dot{M}_0 = 145 \pm 50$ kg s$^{-1}$ ($\dot{M}_0 = 80 \pm 30$ kg s$^{-1}$ if $p_v=0.07$), $t_0$=--2.1 days, FWHM=17.3 days, and $v_0$=47.9 m s$^{-1}$. The total dust mass ejected was then (2.5$\pm$0.9)$\times$10$^8$ kg ( (1.4$\pm$0.5) $\times 10^8$ kg if $p_v=0.07$), all this mass being emitted before the first observation of January 3, 2016. The uncertainties in the dust mass ejected are calculated assuming a criterion for which a fit is not acceptable when $\chi$ exceeds by 10\% of its best-fit value. The maximum radius of the size distribution was finally set to 0.5 cm, as this provides a slightly better fit to the observations than the initially assumed $a_{max}$=1 cm. The best-fit images are compared to the observed ones in Fig. \ref{dust_tail_1}. As it is seen, the simulations reproduce quite accurately the observations at the three epochs. From the synthetic images, the derived apparent magnitudes in the circumnuclear regions (using the same apertures as for the observed images) are compared to the observed ones in Table 2. Both the synthetic as well as the observed magnitudes are referred to r$^\prime$-Sloan. The best-fit value of the synthetic magnitude agrees very well with the observations during both the 2006 and 2016 epochs (see Table 2). 

Taking into account the derived value of $v_0$, the minimum ejection velocity (which occurs for the largest particles of $a_{max}$=0.5 cm) is given by $v_{min}$=0.52 m s$^{-1}$. The escape velocity from the nucleus at a distance of $\sim 20$ $R_N$, where the gas drag vanishes, is given by $v_{esc}$=$\sqrt{\frac{2GM_N}{20 R_N}}$, where $M_N$ is the mass of the nucleus, and $G$ is the universal gravitational constant. Adopting a nucleus density of 1000 kg m$^{-3}$, i.e. assuming a more asteroid-like density for the object \citep{Carr12}, and a radius $R_N \sim 1.15$ km (see next section), we obtain $v_{esc}\simeq$0.19 m s$^{-1}$, so that the largest particles are ejected at velocities well over the escape velocity.  

Having obtained this best fit to the 2016 observations, we now turn our attention to the 2006 apparition images. These observations in principle indicate a similar pattern to the 2016 observations, with the nucleus detached from the tail. We run the model with exactly the same dust parameters found for the 2016 observations. The resulting simulated images are compared to the observations in Fig. \ref{dust_tail_2}, in which scans along the observed and modeled narrow tails are also displayed. As it can be seen, the agreement is very good, confirming that the comet is repeating a very similar behavior during these two apparitions.

\subsection{The nucleus magnitude and size}

\begin{figure}[h]
\resizebox{8cm}{!}{\includegraphics{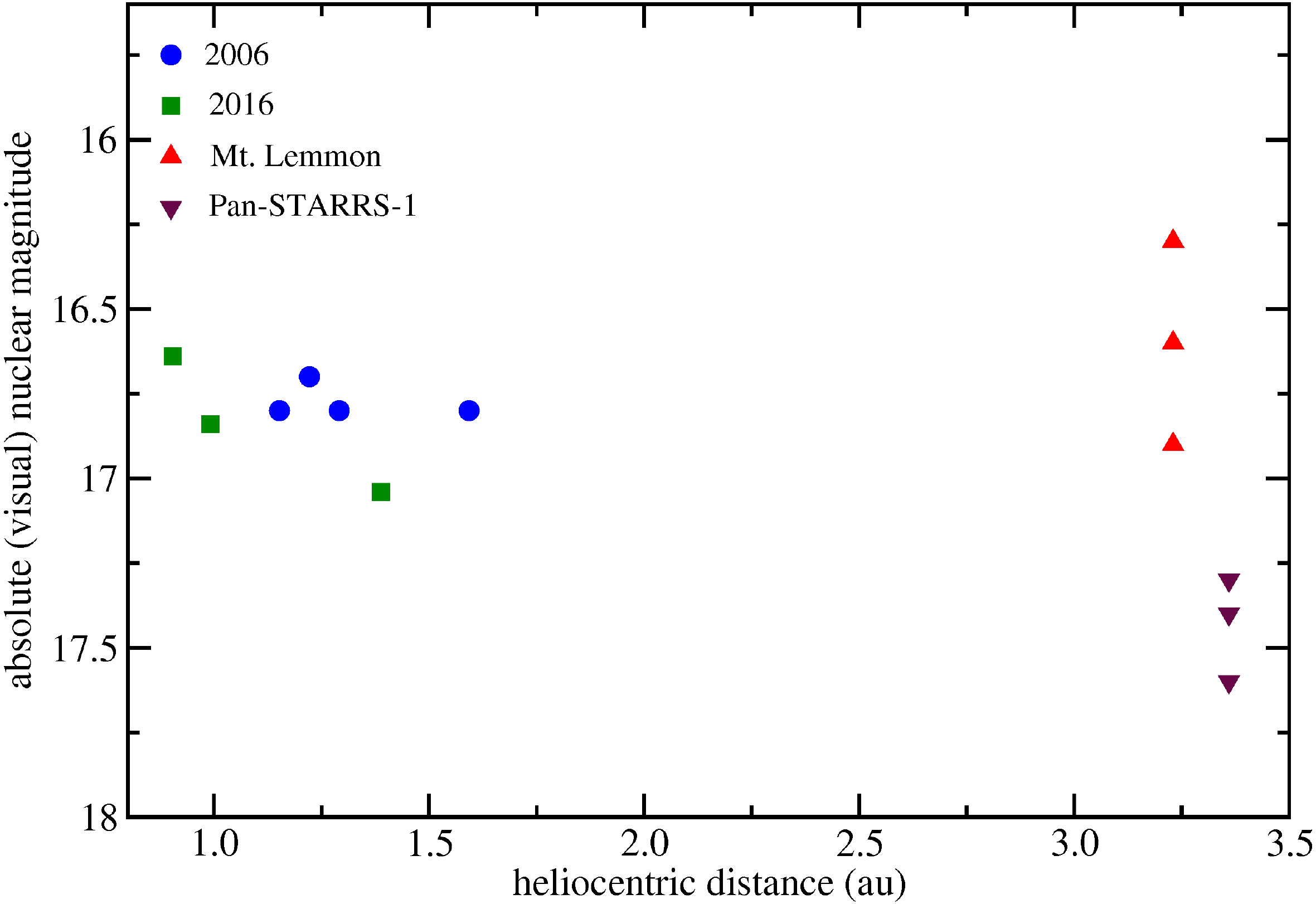}}
\caption{Computed absolute (visual) nuclear magnitude of 249P from the apparent magnitudes shown in Table 2, and from distant observations by the Mt. Lemmon Survey on January 03, 2012 at $r=3.23$ au, and from Pan-STARRS 1 on February 02, 2015 at $r=3.36$ au (Minor Planet Center http://www.minorplanetcenter.net/db\_search).}
\label{magnitudes}
\end{figure}

The apparent magnitudes given in Table 2 seem to correspond to the bare nucleus (or at least with residual activity), except for the image of January 3, 2016, which appears elongated by the comet motion, as the image was obtained using the sidereal tracking, and the comet was still significantly contaminated by the close-by dust tail as discussed in Section 3. We also note that the 2006 and 2016 magnitudes were taken in the SLOAN $r'$ filter. We can convert the $r'$ magnitudes to the more usual Johnson's R and V colors by applying the following color corrections: $R=r'-0.34$ and $V-R=+0.54$ \citep{Cox00}. Therefore, we have $V = r' + 0.2$. From the apparent visual (V) magnitude we can obtain the absolute nuclear magnitude of the nucleus ($H_v$). The results are shown in Fig. \ref{magnitudes} as a function of the heliocentric distances at which they were measured. As said, we discard the January 3, 2016 observation. We also included magnitude estimates at large heliocentric distances by the Mt Lemmon Survey and Pan-STARRS-1. As we can see, there is a reasonable good agreement among the different estimates (within about 0.3-0.4 magnitudes), so we can estimate the visual absolute magnitude of the nucleus at $H_v = 17.0 \pm 0.4$.

If we assume a spherical nucleus of radius $R_N$ and adopt a visual geometric albedo of 0.04 as typical of JFCs, we obtain $R_N=1.32 \pm 0.25$ km. Yet, if this is a peculiar comet of spectral type B, it could well have a somewhat higher albedo. \citet{Alil13}, using WISE data of a large sample of B-type asteroids and a near-Earth thermal model, show that B-types present a narrow albedo distribution with a mean value: $\bar{p}_v =0.07 \pm 0.03$: Thus, for a geometric albedo of 0.07 the radius of 249P would be $R_N = 1.0 \pm 0.20$ km. In the following we will take as our canonical radius $R_N \simeq 1.15 \pm 0.35$ km that takes into account both the uncertainties in the estimated nuclear magnitude and the albedo.      

\section{The post-perihelion decay of the dust production rate. Comparison with some active JFCs}

From the results presented before we find that the dust production rate of 249P peaks near perihelion, and that its activity rapidly decays after an active phase of about 20 days. We can use the photometric data from other observers to check if 249P really became nearly inactive after its brief active phase. To this purpose, we have collected some observed apparent total magnitudes ($m_T$) and coma diameters ($p''$) obtained in the post-perihelion branch. These are shown in Table 3. The sources of these observations are given at the bottom of the table.

\begin{table}[htb]

\centerline{Table 3: Computed dust production rate per unit area ($\xi_d$) of 249P}

\begin{center}

\begin{tabular}{l l l l l l l l} \hline

date & $r$ (au) & $\Delta$ (au) & $\alpha$ & $m_T$ & $p$ ($''$) & $\xi_d$ (kg s$^{-1}$ m$^{-2}$) & Source \\ \hline
2006/10/21 & 1.14 & 0.41 & 59.6 & 17.7 & 8.5 & $1.14 \times 10^{-8}$ & (1) \\
2006/10/21 & 1.14 & 0.41 & 59.6 & 17.6 & 5.0 & $3.59 \times 10^{-8}$ & (1) \\
2015/12/23 & 0.74 & 1.03 & 65.3 & 14.3 & 33.8 & $9.20 \times 10^{-7}$ & (2) \\
2016/01/10 & 1.01 & 1.10 & 55.4 & 15.4 & 33.8 & $3.90 \times 10^{-7}$ & (2) \\
2016/01/17 & 1.11 & 1.18 & 50.8 & 15.3 & 33.8 & $4.50 \times 10^{-7}$ & (2) \\
2016/01/17 & 1.11 & 1.18 & 50.8 & 16.7 & 22.5 & $1.78 \times 10^{-7}$ & (2) \\
\hline
\end{tabular}

\end{center}
(1) IAUC No. 8763\\
(2) {\it Cometas\_Obs Cazadores de Cometas} (http://www.astrosurf.com/cometas-obs/)
\end{table}

A rough estimate of the dust and gas production rate of a comet can be obtained from the knowledge of $m_T$ and the apparent nucleus magnitude $m_N$ at a heliocentric distance $r$. We assume that the coma light comes from scattering of sunlight by dust particles of differential size distribution $n(a)da$. To estimate the dust production rate we will follow in general terms the procedure outlined by \citet{Jewi12}. This procedure essentially applies to low-active comets, where we may assume that there is not much overlapping between dust particles in the coma as seen by the observer. In this case, we can say that the total scattering cross-section of the dust particles in a coma of radius $\Lambda$ is

\begin{equation}
  C_d = \int^{a_{max}}_{a_{min}} Q_s \pi a^2 n(a)da
\end{equation}  
where $Q_s$ is the light scattering efficiency of the dust particles. In the following we assume $Q_s \sim 1$.

The mass of the dust in the coma is

\begin{equation}
  M_d = \int^{a_{max}}_{a_{min}} \frac{4}{3} \pi \rho_d a^3 n(a)da
\end{equation}  
where $\rho_d$ is the mass density of the dust particles.

The weighted mean radius of the dust particles is given by

\begin{equation}
  \bar{a} = \frac{\int^{a_{max}}_{a_{min}} Q_s \pi a^3 n(a)da}{\int^{a_{max}}_{a_{min}} Q_s \pi a^2 n(a)da} = \frac{3M_d}{4\rho_dC_d}
\end{equation}  

If we assume that the coma contains $N_P$ dust particles and that they do not overlap as seen from the observer, we can then express the total optical cross-section of the coma as

\begin{equation}
C_d = N_P\pi\bar{a}^2 = \frac{M_d}{\frac{4}{3}\pi\bar{a}^3\rho_d}\pi\bar{a}^2 =  \frac{M_d}{\frac{4}{3}\bar{a}\rho_d}
\end{equation}

Therefore, solving for $M_d$ we obtain

\begin{equation}
M_d = \frac{4\rho_d\bar{a}C_d}{3}
\end{equation}

  This result is very much dependent on the adopted value of the typical grain size $\bar{a}$ which itself depends on the grain size distribution. For this a power-law size distribution $n(a)da \propto a^{-\kappa}da$ is usually adopted where $\kappa = 3.5$ (cf. Section 4.2). For the upper and lower size limits we can use the values derived in Section 4.2: $a_{min} = 1$ $\mu m$ and $a_{max} = 0.5$ cm. If we integrate $n(a)da$ with these extreme values we obtain $\bar{a} = (a_{min}a_{max})^{1/2} = 7 \times 10^{-3}$ cm.

The dust production rate will be

\begin{equation}
\dot{M}_d = \frac{M_d}{\tau}
\end{equation}  
where $\tau$ is the time that it takes the dust particles to traverse the distance $\Lambda$. $\tau$ depends on the expansion velocity $v_{exp}$ of the dust particles in the coma. To estimate $v_{exp}$ we should bear in mind the expression $v = v_0 \beta^{1/2}$ (cf. Section 4.2) for which we found a best-fit value $v_0 =47.9$ m s$^{-1}$ at about 20 comet radii (cf. Section 4.3). Yet the expansion velocity of the dust particles should increase as they recede from the nucleus, so they may reach terminal velocities of a few hundred m s$^{-1}$ \citep{Liss98,Jewi99}. In the following we will adopt an expansion velocity

\begin{equation}
v_{exp} = v'_0 \beta^{1/2} \sim v'_0 \left(\frac{1.19 \times 10^{-3}}{2\rho_d a}\right)^{1/2} \mbox{ m s$^{-1}$}
\end{equation}
with $v'_0 = 150$ bearing in mind that it may be uncertain by a factor of two. This value may be more representative of comae length scales of $\sim 10^3 - 10^4$ km as those shown in Table 3.

Finally, we have to relate the scattering cross-section of the dust coma with the measured brightness of the coma, $I_C$, within the aperture $p'' \simeq (\Lambda/\Delta) \times 206265$. The brightness of the coma will be compared with that of the bare nucleus, $I_N$. Actually, the total and nuclear magnitudes (for instance in the visual), $m_T$ and $m_N$, are what we usually measure that are related to the brightness by

\begin{equation}
m_T = C -2.5\log{I_T}
\end{equation}  
where the total brightness $I_T = I_d + I_N$.

\begin{equation}
m_N = C -2.5\log{I_N}
\end{equation}  
where $C$ is a constant.

Therefore the ratio of the total to nuclear brightness can be expressed as the difference between the total and nuclear magnitudes, namely

\begin{equation}
\Delta m = m_N - m_T = 2.5\log{\left(\frac{I_T}{I_N}\right)} =  2.5\log{\left(1 + \frac{I_d}{I_N}\right)}
\end{equation}  
Let $C_N = \pi R_N^2$ be the geometric cross-section of the comet nucleus of radius $R_N$. The ratio between the coma brightness and the nucleus brightness will be

\begin{equation}
\frac{I_d}{I_N} = \frac{C_d p_d(\lambda)\phi_d(\alpha)}{C_N p_N(\lambda)\phi_N(\alpha)}
\end{equation}  

In the following we will assume that the albedos and phase functions of dust particles and the comet nucleus are similar, so we can set $p_d(\lambda)\phi_d(\alpha)/p_N(\lambda)\phi_N(\alpha) \simeq 1$. Therefore, substituting eq.(11) into eq.(10), then substituting $C_d$ by the expression given by eq.(4), and finally solving for $M_d$ we get

\begin{equation}
M_d = \frac{4\rho_d\bar{a}}{3}C_d\left(10^{\Delta m /2.5} - 1\right)
\end{equation}  

The dust production rate $Q_d$ is given by

\begin{equation}
Q_d = \frac{M_d}{\tau} = \frac{M_d v_{exp}}{\Lambda}
\end{equation}  
where $\tau \sim \Lambda/v$, and $v_{exp}$ is obtained from eq.(7) with $a=\bar{a}$.

\begin{figure}[h]
\resizebox{8cm}{!}{\includegraphics{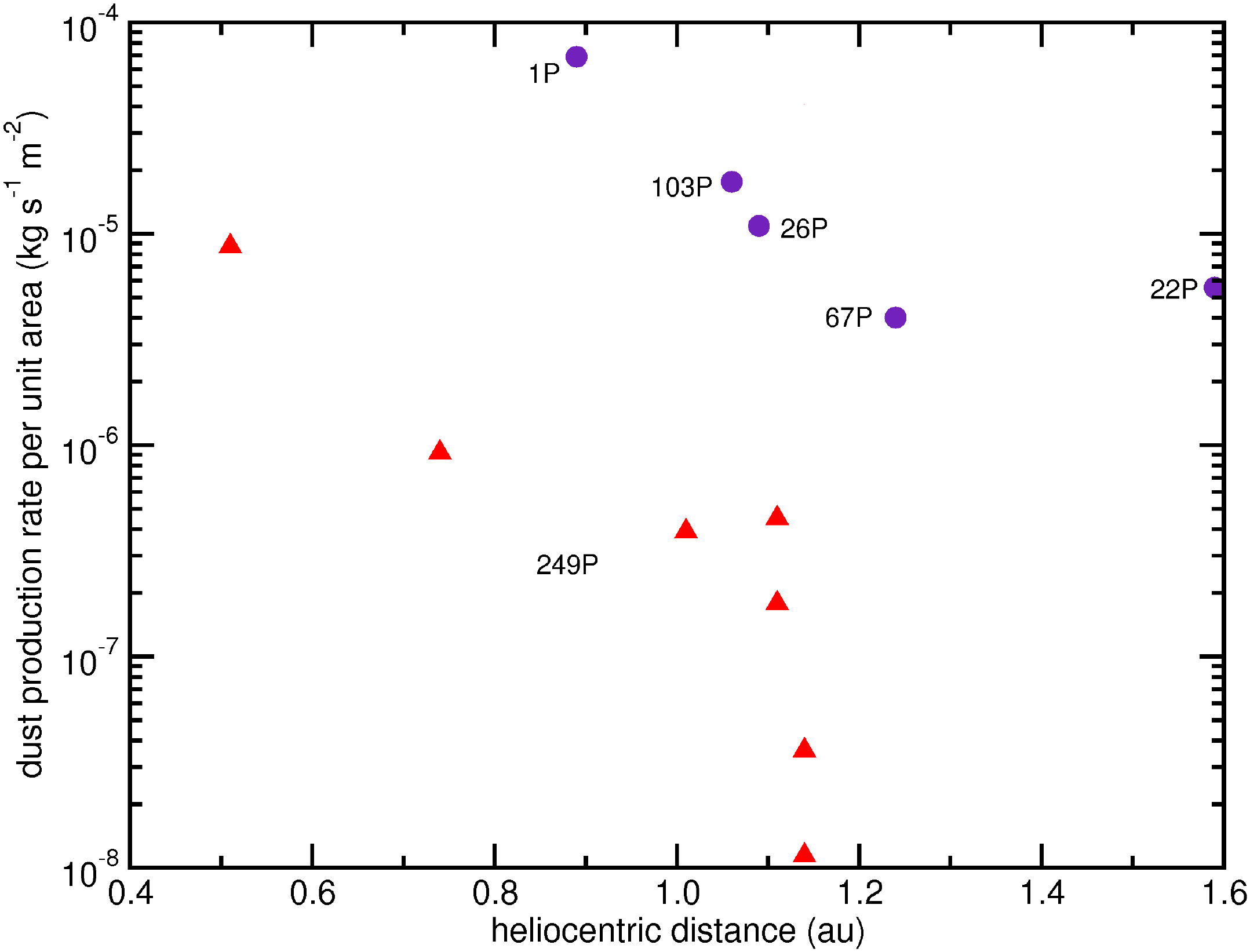}}
\caption{Computed dust production rate per unit area of 249P as a function of the heliocentric distance (red triangles). The value on the left side (close to $r =0.5$ au) was computed by means of the dust tail model of Section 4. The rest of the values were computed by the method developed in this section. As a comparison, we also plot computed $\xi_d$ values for comets 1P/Halley, 22P/Kopff, 26P/Grigg-Skellerup, 67P/Churyumov-Gerasimenko, and 103P/Hartley 2 at the distances at which their dust production rates were measured (violet circles). For the dust production rates we used the following sources : 1P \citep{Thom91}, 22P \citep{Lamy02,Moreno12}, 26P \citep{Full93}, 67P \citep{Ishi08}, 103P \citep{Epif01}. For the sizes we used: 1P \citet{Kell87}, 22P \citet{Lamy02,Fern05}, 26P \citep{Fern05}, 67P \citet{Sier15}, 103P \citet{Thom13}.}
\label{fraction_r}
\end{figure}

If we want to compare the activity of 249P with those of some other comets at similar heliocentric distances, it is more pertinent to define the dust activity per unit area as  

\begin{equation}  
\xi_d = \frac{Q_d}{4\pi R_N^2}
\end{equation}
If we consider that the radius of 249P is $R_N \simeq 1.15 \pm 0.35$ (cf. Section 4.4), the dust production rate per unit area of 249P turns out to be $8.71 \times 10^6$ kg s$^{-1}$ m$^{-2}$ near perihelion, according to our model of Section 4.3. The computed $\xi_d$ in the post-perihelion branch, obtained from the measured photometric magnitudes and coma diameters, are shown in Table 3. As we see, there is a good agreement with our previous assessment of a rapid fall in the dust activity of the comet. In order to compare with the dust production rates observed in other well-known active periodic comets, we show in Fig. \ref{fraction_r} the values of $\xi_d$ for comets 1P/Halley, 22P/Kopff, 26P/Grigg-Skjellerup, 67P/Churyumov-Gerasimenko, and 103P/Hartley 2. The $\xi_d$ values plotted in the figure do not have error bars because of the complexity to estimate all the error sources. Nevertheless we can make a qualitative analysis. For instance, the $\xi_d$ values for 249P may be uncertain within factors of 2-4, taking into account uncertainties in several model parameters and in the nucleus size. As regards the $\xi_d$ values for the active comets, the determination of their dust production is based on observations with error bars (in some cases) of about 20-50\% of the quoted values. As regards, the sizes of our sample of active comets, they are better known than the size of 249P, so we can estimate that $\xi_d$ values for the active comets are correct within a factor of two. Despite these uncertainties, it is clear from Fig. \ref{fraction_r} that the dust production rate per unit area of 249P is one to three orders of magnitude lower than those of active periodic comets at similar heliocentric distances.

\section{Discussion}

The existence of asteroidal NEJFCs raises some questions about their nature and  provenance. First, why can they last for so long ($\gsim 5 \times 10^4$ yr) in the near-Earth region? Second, where is their source region? Do they come from the trans-neptunian region as the rest of the JFCs and by some dynamical process they fall into rather stable orbits or, on the contrary, they come from the main asteroid belt as the NEAs, thus explaining their similar dynamical characteristics? If the latter applies, they could be considered as the counterparts in the near-Earth region of the main-belt comets.

An origin in the main asteroid belt might also imply a more compact and hardened internal structure, reflecting the accretion conditions in the region, in contrast with the icy and fluffy structure of JFCs formed in the trans-neptunian region. The spectroscopic observations showing that 249P has a B-type spectrum -different from the most common D and P types of JFCs- strengthens the view of an origin in the asteroid belt for this peculiar object. We may also presume that 249P has a bulk density higher than the valuesf typical for JFCs (about 500 kg m$^{-3}$, Sierks et al., 2015, and more in line with those found for small asteroids of about 1000 kg m$^{-3}$ or higher, Carry, 2012).
 
The model for the 2016 observations reveals that the comet was significantly active only for $\sim$20 days (FWHM) around perihelion with a modest dust mass loss rate (peak at $145 \pm 50$ kg s$^{-1}$) and an integrated dust production of (2.5$\pm$0.9)$\times$10$^8$ kg for a geometric albedo $p_v =0.04$ (these values slightly decrease when assuming a higher albedo). If this dust production is maintained over 5$\times$10$^4$ years (or 10800 revolutions), the total mass released would be $\sim$2.7 $\times$10$^{12}$ kg, which would correspond to a spherical body of $R_N=0.86$ km with 1000 kg m$^{-3}$ density, namely comparable to the current radius of the comet. Therefore, even though the computed mass loss is very low as compared to other active JFCs, it should be enough to disintergrate the comet over a time scale of 5$\times$10$^4$ yr. There are several possible scenarios to explain the survival of the comet: one of them is that the comet nucleus was larger than it is now 5$\times$10$^4$ yr ago. An alternative explanation is that 249P has now an unusual activity and it used to be much lower in the past. In this regard we note that the perihelion distance of 249P is decreasing to a record low value. In the past it was somewhat more distant to the Sun with perihelion distances $q \sim 0.6-0.8$ au, so it was also presumably less active.

It may also be possible that 249P stays inactive during some time (thus extending its physical lifetime) until a collision with a meteoroid opens a crater and exposes some sub-surface water ice. It is not new the idea that collisions with meteoroids may be a mechanism for new or enhanced activity in comets \citep{Fern81}, and it has also been invoked to explain the activity observed in several main-belt comets \citep{Hagh16}.

It is interesting to stress that most of the activity of 249P was concentrated in a narrow time span near perihelion. What is the reason? It is possible that some combination of orbital motion with the orientation of the spin axis exposes a particular patch of active area or the bottom and lateral walls of an impact crater) to the intense solar radiation near perihelion. As we found, after the active phase of $\sim 20$ days the comet nearly switches off, reducing its activity by one-two orders of magnitude in the following days, that might correspond to the time at which the active patch hides again from the Sun's radiation.

This work opens up a new line of research on the asteroidal NEJFCs. The present study should be extended to other asteroidal NEJFCs (see a list in Fern\'andez and Sosa 2015), whose near-perihelion activity patterns should be scrutinized, besides looking for other interesting features like spectrum, gaseous activity, size and albedo. One of the goals of the study of asteroidal NEJFCs will be to assess if their activity is steady or if it is intermittent through successive perihelion passages, so campaigns to re-observe these objects -starting with 249P- will be of the upmost interest.

\section{Conclusions}

The main conclusions of the work can be summarized as follows:
        
\begin{enumerate}
\item 249P/LINEAR is found to move on an NEA-like dynamically stable orbit. It has been in a near-Earth region for at least $10^4$ yr, and probably for more than $5 \times 10^4$ yr. For most of this time, the comet remained bound in the 5:2 MMR with Jupiter. The long survival near the Sun ($q \simeq 0.4 - 1.1$ au) of a rather small body (radius $\simeq 1.15$ km) may require a mostly rocky, hardened, and more compact structure, as compared to the icy and fluffy structure of typical JFCs. 
\item 249P has been found to be a comet with very little activity despite its very small perihelion distance. The dust production rate at its peak is found to be $145 \pm 50$ kg s$^{-1}$ in its 2016 perihelion passage (for a geometric albedo $p_v = 0.04$). The duration of this level of activity is for about 20 days and then falls off precipitously. This activity pattern is found to be very similar during both the 2016 and 2006 perihelion approaches. The level of activity, as measured by the dust production rate per unit area, is found to be of the order of that found in other active comets only when 249P was close to perihelion. At heliocentric distances $r \sim 1-1.5$ au it fell by one to three orders of magnitude indicating a nearly switch-off of its activity.
\item The spectrum is found to be similar to B-type asteroids, unlike most JFCs that show spectra of types D or P, but that is shared by some main-belt asteroids and in particular some MBCs.
\item The dynamical and physical evidence suggests that we are before an object with a likely different origin from that of typical JFCs that probably come from the trans-neptunian region. By contrast, 249P may be an example of an object coming from the main asteroid belt, thus physically related to the MBCs.
  \end{enumerate}

\vspace{1cm}

{\bf Acknowledgments}

\bigskip

We thank Gian Paolo Tozzi for fruitful discussions. This article is based on observations made with the Gran Telescopio Canarias, operated on the island of La Palma by the Instituto de Astrof\'{\i}sica de Canarias in the Observatorio del Roque de los Muchachos. JAF and AS acknowledge support from the project CSIC Grupo I+D 831725 - Planetary Sciences. JL and JdeL acknowledge support from the project AYA2015-67772-R (MINECO, Spanish Ministry of Economy and Competitiveness). FM acknowledges support from contracts AYA2015-⁠67152-⁠R and AYA2015-⁠71975-⁠REDT from the Spanish Ministerio de Econom\'\i a y Competitividad 

\vspace{1cm}

{\bf References}


\begin{thebibliography}{00}

\bibitem[Alexander et al.(2012)]{Alex12} Alexander, C.M.O'D., Bowden, M., Howard, K.T., Herd, C.D.K., Nittler, L.R., 2012. The provenances of asteroids, and their contributions to the volatile inventories of the terrestrial planets. Science 337, 721-723.

\bibitem[Al\'i-Lagoa et al.(2013)]{Alil13} Al\'i-Lagoa, V., de Le\'on, J., Licandro, J., Delb\'o, M., Campins, H., Kelley, M.S., 2013. Physical properties of B-type asteroids from WISE data. Astron. Astrophys. 554, 71-86.
  
\bibitem[Bus and Binzel(2002a)]{Bus2002a} Bus, S.J., Binzel, R.P., 2002a. Phase II of the small main-belt asteroid spectroscopic survey. The observations. Icarus 158, 106-145.
  
\bibitem[Bus and Binzel(2002b)]{Bus2002b} Bus, S.J., Binzel, R.P., 2002b. Phase II of the small main-belt asteroid spectroscopic survey. A feature-based taxonomy. Icarus 158, 146-177.
  
\bibitem[e.g. Brin(1980)]{Brin80} Brin, G.D., 1980. Three models of dust layers on cometary nuclei. Astrophys. J. 237, 265-279. 

\bibitem[e.g. Carry(2012)]{Carr12} Carry, B., 2012. Density of asteroids. Planet. Space Sci. 73, 98-118.
  
\bibitem[Cepa et al.(2000)]{Cepa2000} Cepa, J., Aguiar, M., Escalera, V. G., et al., 2000. OSIRIS tunable imager and spectrograph. Proc. SPIE, 4008, 623-631.
  
\bibitem[Cepa(2010)]{Cepa2010} Cepa, J., 2010. OSIRIS: Final characterization and scientific capabilities. In: Diego, J.M., Goicoechea, L.J., Gonz\'alez-Serrano, J.I., Gorjas, J. (Eds.), Highlights of Spanish Astrophysics V, Springer-Verlag, Berlin, p. 15.

\bibitem[Chambers(1999)]{Cham99} Chambers, J.E., 1999. A hybrid symplectic integrator that permits close encounters between massive bodies. Mon. Not. R. Astron. Soc. 304, 793-799.  

\bibitem[Cochran and Barker(1987)]{Coch87} Cochran, A.L., Barker, E.S., 1987. Comet Giacobini-Zinner: A normal comet?. Astron. J. 92, 239-243.
  
\bibitem[Cox(2000)]{Cox00} Cox, A.N. 2000, Allen's Astrophysical Quantities, fourth edition. Springer-Verlag.   
  
\bibitem[DeMeo and Binzel(2008)]{Deme08} DeMeo, F., Binzel, R.P., 2008. Comets in the near-Earth object population. Icarus 194, 436-449.

\bibitem[Di Sisto et al.(2009)]{Disi09} Di Sisto, R.P., Fern\'andez, J.A., Brunini, A., 2009. On the population, physical decay and orbital distribution of Jupiter familiy comets: Numerical simulations. Icarus 203, 140-154.

\bibitem[Epifani et al.(2001)]{Epif01} Epifani, E., Colangeli, L., Fulle, M., Brucato, J.R., Bussoletti, E., De Sanctis, M.C., Mennella, V., Palomba, E., Palumbo, P., Rotundi, A., 2001. ISOCAM imaging of comets 103P/Hartley 2 and 2P/Encke. Icarus 149, 339-350.  

\bibitem[Fern\'{a}ndez(1981)]{Fern81} Fern\'andez, J.A., 1981. The role of collisions with interplanetary particles in the physical evolution of comets. Moon Planets. 25, 507-519.
    
\bibitem[Fern\'{a}ndez(2005)]{Fern05} Fern\'andez, J.A. 2005, Comets. Nature, Dynamics, Origin, and their Cosmogonical relevance, Springer, Dordrecht, pp. 240-246.

\bibitem[Fern\'{a}ndez et al.(2014)]{Fern14} Fern\'andez, J.A., Sosa, A., Gallardo, T., Guti\'errez, J.N., 2014. Assessing the physical nature of near-Earth asteroids through their dynamical histories. Icarus 238, 1-12.

\bibitem[Fern\'{a}ndez and Sosa(2015)]{Fern15} Fern\'andez, J.A., Sosa, A., 2015. Jupiter family comets in near-Earth orbits: Are some of them interlopers from the asteroid belt?. Planet. Space Sci. 118, 14-24.

\bibitem[Fern\'{a}ndez(1997)]{YFernandez97} Fern\'andez, Y.R., McFadden, L.A., Lisse, C.M., Helin, E.F., Chamberlin, A.B., 1997. Analysis of POSS images of comet-asteroid transition object 107P/1949 W1 (Wilson-Harrington). Icarus 128, 114-126.

\bibitem[Festou(1981)]{Festou1981} Festou, M.C., 1981. The density distribution of neutral compounds in cometary atmospheres. I - Models and equations. Astron. Astrophys. 95, 69-79.
  
\bibitem[Finson and Probstein(1968)]{Finson68} Finson, M., Probstein, R., 1968, A theory of dust comets. I. Model and equations. Astrophys. J., 154, 327-352.

\bibitem[Fulle(1989)]{Fulle89} Fulle, M., 1989. Evaluation of cometary dust parameters from numerical simulations - Comparison with an analytical approach and the role of anisotropic emissions. Astron. Astrophys. 217, 283-297.

\bibitem[Fulle et al.(2016)]{Full16} Fulle, M., Marzari, F., Della Corte, V., et al., 2016. Evolution of the dust size distribution of comet 67P/Churyumov-Gerasimenko from 2.2 au to perihelion. Astrophys. J. 821:19 (14pp) doi: 10.3847/0004-637X/821/1/19.

\bibitem[Fulle et al.(1993)]{Full93} Fulle, M., Mennella, V., Rotundi, A. Colangeli, L., Bussoletti, E., Pasian, F., 1993. The dust environment of comet P/Grigg-Skjellerup as evidenced from ground-based observations. Astron. Astrophys. 276, 582-588.  

\bibitem[Fukugita et al.(1996)]{Fukugita96} Fukugita, M., Ichikawa, T., Gunn, J.E., et al., 1996. The Sloan Digital Sky Survey Photometric System. Astron. J. 111, 1748-1756.

\bibitem[Gallardo(2006)]{Gall06} Gallardo, T., 2006. Atlas of mean motion resonances in the solar system. Icarus 184, 29-38.  
  
\bibitem[Haghighipour et al.(2016)]{Hagh16} Haghighipour, N., Maindl, T.L., Sch\"afer, C., Speith, R., Dvorak, R., 2016. Triggering sublimation-driven activity of main belt comets. Astrophys. J. 830:22 (11pp).

\bibitem[Harris and D'Abramo(2015)]{Harr15} Harris, A.W., D'Abramo, G., 2015. The population of near-Earth asteroids. Icarus 257, 302-312.   

\bibitem[MBCs e. g. Hsieh and Jewitt(2006)]{Hsie06} Hsieh, H.H., Jewitt, D., 2006. A population of comets in the main asteroid belt. Science 312, 561-563.

\bibitem[Ishiguro(2008)]{Ishi08} Ishiguro, M., 2008. Cometary dust tail associated with Rosetta mission target: 67P/Churyumov-Gerasimenko. Icarus 193, 96-104.

\bibitem[Ishiguro et al.(2015)]{Ishi15} Ishiguro, M., Kuroda, D., Hanayama, H., et al., 2015. Dust from comet 209P/LINEAR during its 2014 return: Parent body of a new meteor shower, the May Camelopardalids. Astrophys. J. Lett. 798:L34 (6pp) doi: 10.1088/2041-8205/798/2/L34.
  
\bibitem[Jewitt(2012)]{Jewi12} Jewitt, D. 2012. The active asteroids. Astron. J. 143, 66-79.

\bibitem[Jewitt and Matthews(1999)]{Jewi99} Jewitt, D., Matthews, H., 1999. Particulate mass loss from comet Hale-Bopp. Astron. J, 117, 1056-1062.  

\bibitem[Keller et al.(1987)]{Kell87} Keller, H.U., Delamere, W.A., Reitsema, H.J., Huebner, W.F., Schmidt, H.U., 1987. Comet P/Halley's nucleus and its activity. Astron. Astrophys. 187, 807-823.
  
\bibitem[Kolokolova et al.(2004)]{kolokolova2004} Kolokolova, L., Hanner, M. S., Levasseur-Regourd, A.-C., Gustafson, B., 2004. Physical properties of cometary dust from light scattering and thermal emission. In: Festou, M.C., Keller, H.U., Weaver, H.A. (eds.), Comets II, Univ. Arizona Press, Tucson, pp. 577-604.

\bibitem[Lamy et al.(2002)]{Lamy02} Lamy, P.L., Toth, I., Jorda, L., Groussin, O., A'Hearn, M.F., Weaver, H.A., 2002. The nucleus of comet 22P/Kopff and its inner coma. Icarus 156, 442-455.
  
\bibitem[Landolt(1992)]{Landolt} Landolt, A., 1992. UBVRI photometric standard stars in the magnitude range 11.5-16.0 around the celestial equator. Astron. J. 104, 340-371.

\bibitem[Licandro et al.(2008)]{Licandro08} Licandro, J., Alvarez-Candal, A., de Le\'on, J., Pinilla-Alonso, N., Lazzaro, D.; Campins, H., 2008. Spectral properties of asteroids in cometary orbits. Astron. Astrophys. 481, 861-877.
  
\bibitem[Licandro et al.(2007)]{Licandroetal2007} Licandro, J., Campins, H., Moth\'e-Diniz, T., Pinilla-Alonso, N., de Le\'on, J., 2007. The nature of comet-asteroid transition object (3200) Phaethon. Astron. Astrophys. 461, 751-757.

\bibitem[Licandro et al.(2011)]{Licandro2011} Licandro, J., Campins, H., Tozzi, G.P., de León, J., Pinilla-Alonso, N., Boehnhardt, H., Hainaut, O.R., 2011. Testing the comet nature of main belt comets. The spectra of 133P/Elst-Pizarro and 176P/LINEAR. Astron. Astrophys. 532, 65-71.
  
\bibitem[Licandro et al.(2013)]{Licandro2013} Licandro, J., Moreno, F., de Le\'on, J., Tozzi, G.P., Lara, L.M., Cabrera-Lavers, 2013. Exploring the nature of new main-belt comets with the 10.4 m GTC telescope: (300163) 2006 VW139. Astron. Astrophys. 550, 17-23.

\bibitem[Lisse et al.(1998)]{Liss98} Lisse, C.M., A'Hearn, M.F., Hauser, M.G., et al., 1998. Infrared observations of comets by COBE. Astrophys. J. 496, 971-991.  
  
\bibitem[Marty(2012)]{Mart12} Marty, B., 2012. The origins and concentrations of water, carbon, nitrogen and noble gases on Earth. Earth Planet. Sci. Letters 313, 56-66.

\bibitem[Mommert et al.(2014)]{Momm14} Mommert, M., Hora, J.L., Harris, A.W., Reach, W.T., Emery, J.P., Thomas, C.A., Mueller, M., Cruikshank, D.P., Trilling, D.E., Delbo, M., Smith, H.A., 2014. The discovery of cometary activity in near-Earth asteroid (3552) Don Quixote. Astrophys. J. 781, 25-34.  
  
\bibitem[Moreno et al.(2012)]{Moreno12} Moreno, F., Pozuelos, F., Aceituno, F., et al., 2012. Comet 22P/Kopff: Dust environment and grain ejection anisotropy from visible and infrared observations. Astrophys. J. 752, 136-147.

\bibitem[Moreno et al.(2016)]{Moreno16} Moreno, F., Snodgrass, C., Hainaut, O., et al., 2016. The dust environment of comet 67P/Churyumov-Gerasimenko from Rosetta OSIRIS and VLT observations in the 4.5 to 2.9 AU heliocentric distance range inbound. Astron. Astrophys. 587, 155-166.  
 
\bibitem[Nelder and Mead(1965)]{Nelder65} Nelder, J.A., Mead, R., 1965. A simplex method for function minimization. Comput. J. 7, 308-313.
  
\bibitem[Press et al.(1992)]{Press92} Press, W.H., Teukolsky, S.A., Vetterling, W.T., Flannery, B.P. 1992, in Numerical Recipes in FORTRAN, Cambridge Univ. Press, 402.  
  
\bibitem[Sierks et al.(2015)]{Sier15} Sierks, H., Barbieri, C., Lamy, P.L., et al., 2015. On the nucleus structure and activity of comet 67P/Churyumov-Gerasimenko. Science 347, DOI: 10.1126/science.aaa1044.

\bibitem[Sitko et al.(2011)]{Sitk11} Sitko, M.L., Lisse, C.M., Kelley, M.S., et al., 2011. Infrared spectroscopy of comet 73P/Schwassmann-Wachmann 3 using the Spitzer Space Telescope. Astron. J. 142:80 (22pp) doi: 10.1088/0004-6256/142/3/80.  

\bibitem[Tancredi et al.(2000)]{Tanc00} Tancredi, G., Fern\'andez, J.A., Rickman, H., Licandro, J., 2000. A catalog of observed nuclear magnitudes of Jupiter family comets. Astron. Astrophys. Suppl. Ser. 146, 73-90.

\bibitem[Thomas and Keller(1991)]{Thom91} Thomas, N., Keller, H.U., 1991. Comet P/Halley's dust production rate at Giotto encounter derived from Halley Multicolour Camera observations. Astron. Astrophys. 249, 258-268.

\bibitem[Thomas et al.(2013)]{Thom13} Thomas, P.C., A'Hearn, M.F., Veverka, J., et al., 2013. Shape, density, and geology of the nucleus of Comet 103P/Hartley 2. Icarus 222, 550-558.

\bibitem[e.g. Weaver et al.(2001)]{Weav01} Weaver, H.A., Sekanina, Z., Toth, I., et al., 2001. HST and VLT investigations of the fragments of comet C/1999 S4 (LINEAR). Science 292, 1329-1333.

\bibitem[Whipple(1951)]{Whipple51} Whipple, F., 1951. A Comet Model. II. Physical relations for comets and meteors. Astrophys. J. 113, 464-474.  
  
\end{thebibliography}

\end{document}